\documentclass[a4paper, 11pt]{article}
\usepackage{amsfonts}
\usepackage{amsmath,amsthm,amssymb,amscd,epsfig}
\usepackage{graphicx}
\usepackage{psfrag}
\usepackage{amsmath}

\newcommand {\N}{\mathbb{N}}

\newcommand {\rd}{\mathrm{d}}

\numberwithin{equation}{section}

\begin{document}

\title{Equilibration, generalized equipartition, and diffusion in dynamical Lorentz gases}
\author{S. {\ De Bi\`{e}vre}\thanks{Stephan.De-Bievre@math.univ-lille1.fr} \\
Laboratoire Paul Painlev\'e, CNRS, UMR 8524 et UFR de Math\'ematiques \\
Universit\'e Lille 1, Sciences et Technologies\\
F-59655 Villeneuve d'Ascq Cedex, France.\\
Equipe-Projet SIMPAF \\
Centre de Recherche INRIA Futurs \\
Parc Scientifique de la Haute Borne, 40, avenue Halley B.P. 70478 \\
F-59658 Villeneuve d'Ascq cedex, France.\\
and\\
P.~E. {Parris}\thanks{%
parris@mst.edu} \\
Department of Physics\\
Missouri University of Science \& Technology,\\
Rolla, MO 65409, USA\\
}
\date{\today}
\maketitle

\begin{abstract}
We prove approach to thermal equilibrium for the fully Hamiltonian dynamics
of a dynamical Lorentz gas, by which we mean an ensemble of
particles moving through a $d$-dimensional array of fixed
soft scatterers that each possess an internal harmonic or anharmonic degree
of freedom to which moving particles locally couple. We establish that the
momentum distribution of the moving particles approaches a Maxwell-Boltzmann
distribution at a certain temperature $T$, provided that they
are initially fast and the scatterers are in a sufficiently
energetic but otherwise arbitrary stationary state
of their free dynamics--they need not be in a state of
thermal equilibrium. The temperature $T$ to which the particles
equilibrate obeys a generalized equipartition relation,
in which the associated thermal energy $k_{\mathrm B}T$ is equal
to an appropriately defined average of
the scatterers' kinetic energy. In the equilibrated state,
particle motion is diffusive.
\end{abstract}

\section{Introduction}

While it is a well established fact that large systems converge to thermal equilibrium starting
from a more or less arbitrary initial state,
few or no general \emph{dynamical} mechanisms responsible
for this approach to equilibrium have been identified.
An exception is the case where a small system $S$ with Hamiltonian $H_S$ and only a few degrees of
freedom is weakly coupled to a large ``reservoir'' $R$ with Hamiltonian $H_R$ and many degrees of freedom.
Under these circumstances, the small system will converge from an arbitrary initial state $\rho_0$ to a
state $\rho_S\sim\exp(- H_S/k_{\mathrm B}T)$ of thermal equilibrium provided that the reservoir is itself initially in a
thermal state $\rho_R\sim\exp(-H_R/k_{\mathrm B}T)$.
This phenomena of \emph{return to equilibrium} has been proven rigourously in a number of (relatively simple) systems~\cite{jp, bfs, dj}. In such
systems return to equilibrium occurs because, roughly speaking,
the smaller subsystem acts as a small perturbation of the larger one. Strong
stability properties of the thermal states $\rho\sim\exp(-H/k_{\mathrm B}T)$ then force the
coupled system to a joint equilibrium state characterized by the initial
temperature of the reservoir. Studies of this kind however, do not address the question of how the past evolution of the reservoir dynamically led to its degrees of freedom being in thermal equilibrium in the first place.

In this paper we demonstrate \emph{approach to equilibrium} within a class of fully Hamiltonian models that we refer to as dynamical Lorentz gases.  In these models, each member of an ensemble of particles moves
through an array of localized independent scatterers, each possessing an internal harmonic or anharmonic degree
of freedom to which the particle locally couples [see~\eqref{eq:eqmotion}].  The degrees of freedom of the scatterers are initially drawn from a \emph{stationary but not necessarily thermal state} of their uncoupled dynamics.
Our main result is that, asymptotically in time, as a result of repeated scattering events between a
particle and the degrees of freedom of the medium, the particle momentum distribution $\rho_t(p)$ will,
starting from an arbitrary initial distribution $\rho_0(p)$ of sufficiently large
average mean speed, inevitably converge to a Maxwell-Boltzmann distribution characterized by a
well-defined, non-zero temperature $T$. This constitutes a true approach (rather than return) to
equilibrium since this emergence of the Maxwell-Boltzmann distribution
occurs even when the scattering medium, which  serves as a reservoir
for the particle, is not itself in a state of thermal equilibrium to
begin with. Our result therefore helps to explain, from a purely dynamical point of view,
the robustness and ubiquity of the Boltzmann factor that characterizes states of
thermal equilibrium.

In our analysis, the effective temperature $T$ that
dynamically emerges from the interaction between the moving particle and the
scatterers leads to a \emph{generalized equipartition}
relation [see~(\ref{eq:Tweighteddav})], in which the thermal
energy $k_{\mathrm B}T/2$ per degree of freedom of the particle is equal to a
suitably-defined coupling-weighted average of
the kinetic energy of the scatterers.
In two particular cases it reduces  to the standard
(i.e., un-weighted) equipartition result: (i) when the
coupling of the particle to
a scatterer is linear in
the displacement from mechanical equilibrium of the latter's coordinate
(independent of the particular stationary
distribution of the scatterers), and (ii) when the scatterers are themselves
initially in thermal equilibrium at a well-defined temperature $T$
(independent of the specific coupling between the scatterers and the
particle).
As an example of a situation in which the standard equipartition result
is not obeyed, we
analytically predict, using the generalized equipartition relation, and 
confirm numerically, 
that when the particle is
quadratically coupled to ``harmonic scatterers'' that are each initially
in their own micro-canonical state of fixed energy,
the final thermal energy of the
particle is exactly
one-half the value expected from standard equipartition.

We show finally that in these models the asymptotic mean-squared
displacement of an initially localized ensemble of moving particles grows
linearly in time, with a diffusion constant $D$ that depends on
the effective temperature $T$ through the relation [see~\eqref{eq:DT}]
$$
D\sim T^\nu, \quad\mathrm{with}\quad \frac12\leq\nu< \frac52,
$$
where the power $\nu$ depends on the nonlinearity of the coupling
and on the anharmonicity of
the scatterers. A more strongly nonlinear
coupling leads to a lower value of $\nu$, and to slower
diffusion. Stronger scatterer anharmonicity, on the other hand, leads to
a higher value of $\nu$, and to faster diffusion.

Our demonstration of these results depends upon a careful analysis of the
interaction between the moving particles and the local scatterers, and
notably of the energy exchange that occurs between them. Based on the
Hamiltonian nature of the dynamics we show that, for sufficiently fast
particles, although random fluctuations in the energy change of the particle
(per scattering event) have a magnitude that is of order $\Vert p\Vert^{-1}$, the \emph{average}
energy change is \emph{negative} at high energies, and of order $\Vert p\Vert^{-2}$ [see~(\ref{eq:deltaEndominant}),~(\ref{eq:beta1averages}) and~(\ref{eq:beta2average})]. Thus, while
random fluctuations tend to induce a diffusive growth in the particle's
speed to ever higher values, a weaker (but systematic) average
energy \emph{loss} acts as a
source of \emph{dynamical friction}~\cite{chandraI} that
tends to reduce it. The
relative strengths of these two processes, as encoded in the precise power
laws above, are of the exact form required to dynamically drive the kinetic
energy distribution of the particle asymptotically in time to a Boltzmann distribution, with a
temperature $T$ determined by the ratio of the magnitude of the
fluctuations to that of the dynamical friction [see~(\ref{eq:betastar})]. We stress that,
in
our analysis, this fluctuation-dissipation-like relation emerges naturally
from the microscopic Hamiltonian dynamics, and acts as a defining property of the temperature of
the limiting Boltzmann distribution,
rather than as an a posteriori property of thermal equilibrium, as it does in
most treatments (see for example~\cite{kuboII}).

The results presented here constitute a generalization in various ways of
our previous work \cite{spd} in collaboration with A.~Silvius. There,
diffusive behavior was proven for a particle moving through a
one-dimensional lattice of harmonic scatterers to which the particle was
coupled linearly with a very particular form factor. Approach to
equilibrium, although observed numerically, was not addressed in that paper.
Our work here benefits from the insights gained since in \cite{adblafp},
where the related but very different problem of the motion of fast particles
in a random time-dependent potential was studied. The potentials considered
in that paper, while also generated by an array of isolated time-evolving
scatterers, are non-reactive
(or inert), in that they do not respond to the particle's
passage. The total energy of such a system is not conserved, and the random
scattering events experienced by the particle cause its kinetic energy to
grow slowly but indefinitely, in the phenomenon
of \emph{stochastic acceleration}.
As discussed above, however, when the scatterer degrees of freedom are
treated dynamically, as they are in the current paper, the Hamiltonian
interaction between the scatterers and the moving particle provides a
way for the particle to dissipate excess energy to the
medium; this mechanism completely suppresses stochastic acceleration
and allows the particle to approach thermal equilibrium. In fact, in the
Hamiltonian models
considered here, the frictional component of the force cannot be independently made
small compared to the strength of its fluctuating part. As a result, in such systems
stochastic acceleration cannot be observed, even on intermediate time-scales,
before equilibration sets in.

\begin{figure}[ptb]
\begin{center}
\includegraphics[height=6cm,keepaspectratio=true]{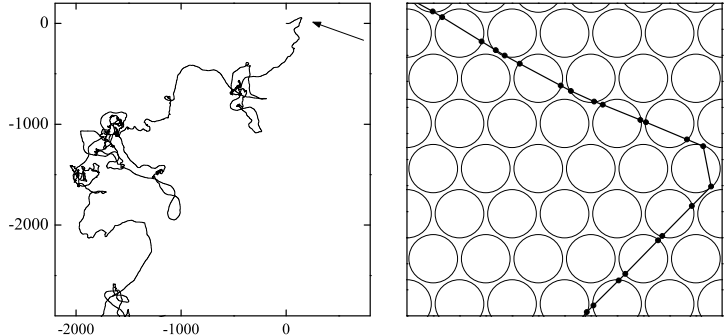}
\end{center}
\caption{The left panel shows an extended part of a typical trajectory, starting
from the origin, computed for a particle moving through a hexagonal lattice of
harmonic scatterers, with unit lattice spacing and linear coupling,
as described more fully in Sec.~\ref{s:numerics}.
The right panel
depicts a close-up view of the region indicated by an arrow in the left panel. In
the close-up, the points where the particle enters and exits each circular interaction
region are indicated with black dots.}
\label{fig:typicaltraj}
\end{figure}

We now describe more precisely the models that form the focus of our
analysis. After a dimensional re-scaling of the dynamical variables, masses,
coupling constants, and time into appropriate dimensionless forms, the
particle-scatterer system in our model is assumed to obey the following equations of
motion
\begin{equation}\label{eq:eqmotion}
\left.
\begin{split}
\ddot{y}(t) & =-\alpha \sum_{N}\eta(Q_{N}(t))\nabla\sigma(y(t)-x_{N}) \\
M\ddot{Q}_{N}(t)+ U'(Q_{N}(t)) & =-\tilde{\alpha}\eta'
(Q_{N}(t))\sigma(y(t)-x_{N}).
\end{split}
\right\}
\end{equation}
In these equations\footnote{All variables and constants appearing in this equation of motion can be
thought of as being dimensionless. They can be obtained from a dimensional
model
\begin{align*}
\hat{m}\frac{d^{2}\hat{y}}{d\tau^{2}} & =-\hat\alpha\sum_N\hat{\eta }(%
\hat{Q}_{N}(\tau)/\ell)\nabla\sigma((\hat{y}(t)-\hat{x}_{N})/L) \\
\hat{M}\frac{d^{2}\ddot{Q}_{N}}{d\tau^{2}}+\hat{k} U'(\hat{Q}%
_{N}(\tau)/\ell) & =-\hat{\tilde\alpha}\eta'(\hat{Q}%
_{N}(\tau)/\ell)\sigma((\hat{y}(t)-\hat{x}_{N})/L),
\end{align*}
where $\ell$ and $L$ are lengths, $\hat{m}$ and $\hat{M}$ masses, $%
\hat k,\hat\alpha,\hat{\tilde\alpha}$ energies, by introducing
\begin{equation*}
\omega^{2}=\frac{\hat{k}}{\hat{m}\ell^{2}},\quad M=\frac{\hat{M}L^{2}}{%
\hat{m}\ell^{2}},\quad\alpha=\frac{\hat\alpha}{\hat{m}\omega
^{2}\ell^{2}},\quad\tilde{\alpha}=\frac{\hat{\tilde\alpha}}{\hat {m}%
\omega^{2}\ell^{2}}
\end{equation*}
and performing the change of variables
\begin{equation*}
t=\omega\tau,\quad y(t)=\hat{y}(\tau)/L,\quad Q(t)=\hat{Q}(\tau)/\ell,\quad x_N=\hat x_N/L,
\end{equation*}
which leaves a dimensionless model governed by three independent
dimensionless parameters, $M,\alpha,$ and $\tilde{\alpha}$.}, 
$y(t)\in\mathbb{R}^{d},$ $d>1$ represents the particle position at
time $t,$ and $Q_{N}\in\mathbb{R}$ the displacement of the internal degree
of freedom associated with the scatterer centered at the fixed point $%
x_{N}\in\mathbb{R}^{d}$. The potential $U$ governing the uncoupled
dynamics of the scatterer (i.e., of its internal degree of freedom)
as well as the coupling function $\eta$ are
assumed to be  smooth functions of their argument. More precisely,
we will always assume that $U$ and $\eta$ exhibit polynomial growth of the type $
U(Q)\sim|Q|^{r}$, $|\eta(Q)|\sim |Q|^{r'}$, for some integers $0<r'\leq r$, so
that in particular $U$ is confining. The simplest case, in which $\eta$ is a linear
function of its argument will be referred to as ``linear coupling''.
The locations $x_{N},N\in\mathbb{Z}^{d}$ of the scattering centers can be
chosen either randomly (with uniform density) or lying on a regular lattice.
The form factor $\sigma(\cdot)$ appearing in the interaction terms is
assumed to be a rotationally invariant smooth function, bounded by one in
absolute value and supported in a ball of radius $1/2$, so that the particle
interacts with the scatterer at $x_{N}$ only when $\Vert
y(t)-x_{N}\Vert\leq1/2$. We suppose $\min_{N\not =M}\Vert
x_{N}-x_{M}\Vert>1$, so that the interaction regions associated with
different scatterers do not overlap, and further assume that the system has
a finite horizon, so that the distance over which a particle can freely
travel without encountering a scatterer, is less than some fixed distance $%
L_{\ast}>0$, uniformly in time and space and independently of the direction
in which it moves.
Finally, $\alpha,\tilde{\alpha}\geq0$ are coupling constants that we assume to
be small. A typical
numerically computed trajectory $y(t)$ is presented in Fig.~\ref{fig:typicaltraj}, for
a system of the above type,
described more fully in Sec.~\ref{s:numerics}, in which
the scatterers are harmonic oscillators, the coupling is linear, and
the form factor $\sigma$ is equal to unity inside the circular interaction
regions indicated, and vanishes everywhere else.

In the inert models
studied in \cite{adblafp}, $\tilde{\alpha}=0$ so that the scatterers do not
respond to the presence of the particle [see~\eqref{eq:eqmotion}]. In this paper we are interested in
the case where they do, and take $\tilde{\alpha}=\alpha$. In this
case the equations (\ref{eq:eqmotion}) are, with $p=\dot{y}$ and $P_{N}=M\dot{Q}_N$, equivalent to
the energy conserving equations of motion generated by
the Hamiltonian
\begin{equation*}
H=\frac{p^{2}}{2}+\sum_{N}H_{\mathrm{scat}}(Q_{N},P_{N})+\alpha\sum_{N}%
\eta(Q_{N})\sigma(y-x_{N}),
\end{equation*}
where
\begin{equation}
H_{\mathrm{scat}}(Q,P)=\frac{P^{2}}{2M}+U(Q). \label{eq:H0}
\end{equation}
Obviously, it is not strictly necessary in our analysis to introduce the
separate coupling parameter $\tilde{\alpha}$. However, as we will see,
retention of this extra parameter in the analysis helps to clearly bring out
the source of the dynamical friction exerted on the particle by
the scatterers, and to aid in its identification with the back-reaction of
the scatterer.

It should also be obvious that, in this class of models, evolution in time
of the internal degree of freedom $Q_{N}\left( t\right) $ of a scatterer
does not lead to a corresponding change in its \emph{location} $x_{N}$, nor
in the size or the shape of the interaction region experienced by a moving
particle that encounters it. It does, however, lead to a time-dependent
change in the magnitude (and possibly the sign) of the interaction energy
between the particle and the scatterer during the time that the particle
traverses the interaction region associated with it. When $\tilde{\alpha}=0$
and $\alpha \eta=+\infty,$ the scatterers form isolated and impenetrable
hard spherical obstacles; in this limit the model reduces to a standard
Lorentz gas. In our generalization of such a gas, $\alpha\eta$ is assumed to
be finite, so the scatterers are soft; a particle with high enough energy
can then enter the interaction region associated with each one. Moreover, as
noted above, with $\tilde{\alpha }=\alpha,$ the internal degree of freedom
of each scatterer responds dynamically to the presence of the moving particle
in an energy-conserving manner, providing a mechanism for the moving
particle to dissipate excess energy to the scatterer degrees of freedom. It
is these features, taken together, that motivate our reference to this class
of models as dynamical Lorentz gases.

The rest of the paper is organized as follows. In Sec.~\ref{s:prw} we
develop a coupled random walk description of the evolution of the moving
particle's momentum and position that forms the foundation upon
which the rest of our analysis is
based. In Sec.~\ref{s:equilibration} we show how this leads to an uncoupled
random walk in the particle's energy, which we use to obtain a Fokker-Planck
equation for the particle's speed distribution function. This allows us to
prove our main result regarding the approach of the momentum distribution to
equilibrium. In Sec.~\ref{s:numerics} we further explore the
implications of our analytical result for
the final particle temperature, and use the result of that analysis
to make predictions for specific dynamical models. We
then present numerical results, based on computations of the sort appearing
in Fig.~\ref{fig:typicaltraj}, that confirm the basic predictions of our analysis.
In Sec.~\ref{s:diffusion} we consider the particle's motion in
position space, show it to be diffusive, and calculate the dependence of the
diffusion constant on the effective temperature $T$ to which the
particle equilibrates.
We also discuss how the stochastic acceleration of the particle,
present when $\tilde\alpha=0$, is suppressed
in a reactive medium, with $\tilde\alpha=\alpha$.
Additional numerical results supporting our calculation
of the diffusion constant are also presented in that section.
Section~\ref{s:discussion} contains a summary, and a discussion
of the connection of our results to those appearing in early
work of Chandresekhar~\cite{chandraI,chandraII,chandraIII}.
The Appendix contains details of the perturbative expansions that underly
our analysis of Sec.~\ref{s:equilibration}.

\section{Particle in a field of scatters: a random walk description}

\label{s:prw} We first recall the random walk description
developed in \cite{adblafp} to describe a particle moving through an array
of random but non-reacting ($\tilde{\alpha}=0$) scatterers,
and generalize it to the present
situation in which the internal degree of freedom of the scatterer responds
dynamically to the presence of the moving particle~($\tilde\alpha=\alpha$). To that end, we consider
a typical trajectory generated by the equations of motion (\ref{eq:eqmotion}%
) for a particle that successively encounters scattering
centers $y_{n} := x_{N_{n}}$
at a sequence of instants $t_{n}$, with incoming momenta $p_{n}$ and impact
parameters $b_{n}$. These quantities are related by (see Fig.~\ref%
{fig:collision})
\begin{equation*}
y_{n}^{-}:=y\left( t_{n}\right) =x_{N_{n}}-\frac{1}{2}e_{n}+b_{n},\quad
e_{n}=\frac{p_{n}}{\Vert p_{n}\Vert}\quad b_{n}\cdot e_{n}=0,\quad\Vert
b_{n}\Vert\leq1/2.
\end{equation*}
\begin{figure}[ptb]
\begin{center}
\includegraphics[height=7cm,keepaspectratio=true]{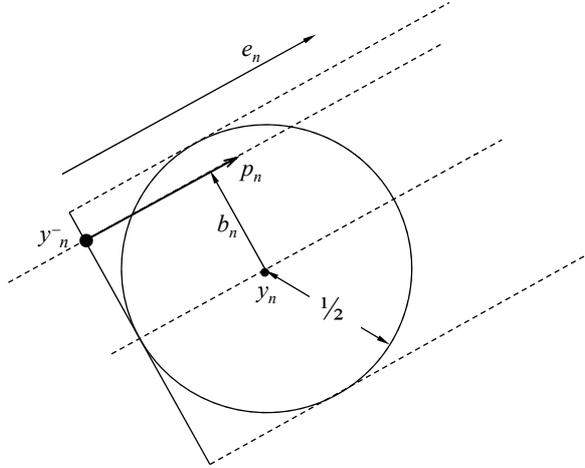}
\end{center}
\caption{A particle at time $t_{n}$ impinging with momentum $p_{n}$ and
impact parameter $b_{n}$ on the $n$th scatterer, centered at the point $%
y_{n} = x_{N_{n}}.$}
\label{fig:collision}
\end{figure}
Note that we view the impact parameter $b_{n}$ as a vector perpendicular to
the incoming direction. We choose $t_{0}=0$ and $x_{N_{0}}=0$, with the
particle starting near a scatterer located at the origin. The initial
displacements and momenta of the scatterers are assumed to be
identically and independently distributed according to a stationary state
\begin{equation}
\rho(Q,P)=\hat{\rho}(H_{\mathrm{scat}}(Q,P))=\rho(Q,-P) \label{eq:rhocond}
\end{equation}
of the scatterer Hamiltonian (\ref{eq:H0}).
At each scattering event the momentum of the particle undergoes a change $%
\Delta p_{n}=p_{n+1}-p_{n}$. Since no forces act on the particle in the
region between scattering centers, the momentum $p_{n+1}$ with which it
leaves the $n$-th scatterer is also the one with which it impinges on the
next. The momentum change $\Delta p_{n}$ depends on the impact parameter $%
b_{n}$ and on the arrival time $t_{n}$ at the $n$th scattering center,
through the displacement $Q_{n}:=Q_{N_{n}}(t_{n})$ and momentum $%
P_{n}:=P_{N_{n}}(t_{n})$ of that scatterer at the time the particle
encounters it. Specifically, we can write%
\begin{equation}
\Delta p_{n}=R(p_{n},b_{n},Q_{n},P_{n}),  \label{eq:deltap}
\end{equation}
where
\begin{equation}
R(p,b,Q,P)=-\alpha\int_{0}^{+\infty}\mathrm{d}\tau\ \eta(Q(\tau))\nabla
\sigma(q(\tau))  \label{eq:R}
\end{equation}
in which $q(\tau),Q(\tau)$ (i.e., without subscripts) denote the unique
solution of the \emph{single scatterer problem}
\begin{equation}\label{eq:onehitequation}
\left.
\begin{split}
\ddot{q}(t)= & -\alpha\eta(Q(t))\nabla\sigma(q(t)),   \\
M\ddot{Q}(t)= & - U'(Q(t))-\tilde{\alpha}\sigma(q(t))\eta'(Q(t)),
\end{split}
\right\}
\end{equation}
with initial conditions
\begin{equation}\label{eq:inicond}
q(0)=b-\frac{1}{2}p/\Vert p\Vert,\quad p(0)=p,\quad Q(0)=Q
\quad\mathrm{and}\quad P(0)=P.
\end{equation}

After leaving the support of the $n$th scatterer, the particle travels a
distance $\ell_{n}$ to the next, which it reaches after a time $\Delta
t_{n}=t_{n+1}-t_{n}=\ell_{n}/\Vert p_{n+1}\Vert$ with impact parameter $%
b_{n+1}$, that both depend on the geometry of the array of scatterers, and
on the dynamics of the $n$th scattering event through the precise point $%
y_{n}^{+}=y_{n}\left( t_{n}^{+}\right) $ from which the particle leaves the $%
n$-th scatterer, and through the outgoing direction $e_{n+1}$. The particle
finds the next scatterer in the state $(Q_{n+1},P_{n+1})$. The process then
repeats itself.

Based upon this description of the dynamics, and ignoring the role of
recollisions, we argue, as in~\cite{adblafp}, that the motion of an ensemble
of particles through an array of scatterers of this type is well
approximated by a coupled discrete-time random walk in momentum and position
space. Each step of the walk corresponds to one scattering event, where the
variables $(Q_{n},P_{n})$ that characterize the scatterer, and the variables
${\ell}_{n}{,b}_{n}$ that characterize the approach of the particle onto the
scatterer, are drawn from the distributions that govern them in the actual
system of interest. In this manner, starting from a given initial condition $%
\left( q_{0},p_{0}\right) $, the velocity $p_{n+1}$, the location $y_{n+1}$,
and the time $t_{n+1}$ of the particle immediately before scattering event $%
n+1$ are iteratively determined through the relations
\begin{equation}
\left.
\begin{array}{lll}
p_{n+1} & = & p_{n}+R\left( p_{n},\kappa_{n}\right) \\
t_{n+1} & = & t_{n}+\frac{\ell_{\ast}}{\Vert p_{n+1}\Vert} \\
y_{n+1} & = & y_{n}+\ell_{\ast}e_{n+1}%
\end{array}
\right\}  \label{eq:finalrw}
\end{equation}
where $\kappa_{n}=(b_{n},Q_{n},P_{n})$ and where $R\left( p,\kappa\right) $ is
defined by (\ref{eq:R}) and (\ref{eq:onehitequation})-(\ref{eq:inicond}). In
this process, the
parameters $(Q_{n},P_{n})$ are assumed to be independently and identically
chosen from the assumed stationary distribution $\hat{\rho}\left( H_{\mathrm{%
scat}}\left( Q,P\right) \right) $, while the $b_{n}$ are independently
chosen at each step uniformly from the $d-1$ dimensional ball of radius $1/2$
perpendicular to $p_{n}$. Without the loss of any essential physics, we have
in (\ref{eq:finalrw}) replaced the random variable $\ell_{n}$ at each time
step with the average distance $\ell_{\ast}<L_{\ast}$ between scattering
events. As a result (\ref{eq:finalrw}) defines a Markovian random walk
determined completely by $p_{0}$ and the random process $(\kappa _{n})_{n\in%
\mathbb{N}}$. Note that, for each $n$, when $p_{n}$ is defined by (\ref%
{eq:finalrw}), $\kappa_{n}$ is independent of $p_{n}$, since the latter only
depends on values of $\kappa_{k}$ with $k<n$. In what follows we write $%
\left\langle \cdot\right\rangle $ for averages over all realizations of the
random process $\kappa_{n}$. In addition, for a function $f$ depending on $p$
and $\kappa=(b,Q,P)$, $b\cdot p=0,\Vert b\Vert\leq1/2$ we denote the average
over $\kappa$ by
\begin{equation}
\overline{f\left( p\right) }=\int_{\Vert b\Vert\leq1/2}\frac{\mathrm{d}b}{%
C_{d}}\int\ \mathrm{d}Q\ \mathrm{d}P\ \rho(Q,P)f(p,b,Q,P),
\label{eq:overlinedef}
\end{equation}
where $C_{d}$ is the volume of the ball of 
radius $1/2$ in $\mathbb{R}^{d-1}$.

\section{Equilibration}\label{s:equilibration}
In this section, we show that, for suitable initial conditions described below,
the distribution of
particle speeds for an ensemble of particles evolving according to the
random walk ({\ref{eq:finalrw})} asymptotically approaches a Maxwellian
distribution, and derive expressions for the temperature $T$ to which
it equilibrates. To this end, we note that the first equation defining the
coupled random walk in (\ref{eq:finalrw}) is independent of the remaining
two. This allows us to temporarily ignore the spatial motion of the
particle, and the time, and independently focus on the random walk that
occurs in the momentum $p_n$ and/or the kinetic energy $E_n =\|p_n\|^2/2$ of the
particle as a function of the step (i.e., collision) number $n$. Thus, for
example, the energy exchanged
\begin{equation}
\Delta E\left( p,\kappa\right) =\frac{1}{2}\left( \left( p+R\left(
p,\kappa\right) \right) ^{2}-p^{2}\right)  \notag
\end{equation}
during a single collision between a particle of fixed incoming momentum $p$
and a scatterer, can be written in terms of the solutions to (\ref%
{eq:onehitequation}) as%
\begin{equation}
\Delta E=\alpha\int_{0}^{t_{+}}\Lambda(s)\sigma(q(s))\mathrm{d}s\quad\mathrm{where}\quad \Lambda(s)=\eta^{\prime}(Q(s))\dot{Q}(s),
\label{eq:deltadef}
\end{equation}
and  where $t_{+}$ is the instant of time
at which the particle emerges from the interaction region ($\|q\|\leq 1/2$)
at the end of the scattering event; note that it can be replaced in the integral by $%
+\infty$.

In the following we will suppose that the scatterer distribution $\rho$ has a
finite mean energy
\begin{equation*}
 E_*=\int\ H_{\mathrm{scat}}(Q,P)\hat\rho(H_{\mathrm{scat}}(Q,P))\rd Q\rd P
\end{equation*}
and that the probability of finding a scatterer with an energy much higher than $E_*$ is negligibly small.
Both the normalized Liouville measure on the energy surface $H_{\mathrm{scat}}(Q,P)=E_*$, associated with the
 microcanonical distribution, and the Boltzmann distribution
$Z_\beta^{-1}\exp(- H_{\mathrm{scat}}/k_{\mathrm B }T)$, for $T$ large enough, satisfy this condition. In addition, we will assume that the particles are almost always both energetic and fast. By energetic, we mean that the particle has a kinetic energy $\|p_n\|^2/2$ well above the typical interaction potential $\alpha\eta(Q_{N_n})\sigma(y(t_n)-x_{N_n})$ that it encounters
in any scattering event. Since $\eta(Q)\sim Q^{r'}$, and $U(Q)\sim |Q|^r$, with $r'\leq r$, the typical size of
$\eta (Q)$ is of order $E_*^{r'/r}$ at most, so this first condition
can be expressed as
\begin{equation}\label{eq:energetic}
 \|p_n\|^2>>\alpha E_*^{r'/r}.
\end{equation}
In addition, we need the particles to be fast, which means they cross the
interaction region in a time $\|p\|^{-1}$ that is short compared to the typical time over which the scatterer evolves. Since the period $\tau_{E_*}$ of $H_{\mathrm{scat}}$ at energy $E_*$ behaves as
$
\tau_{E_*}\sim E_*^{\frac1r-\frac12},
$
this second requirement is equivalent to the relation
\begin{equation}\label{eq:fast}
\|p_n\|^2>> E_*^{1-\frac2r}.
\end{equation}

The conditions~(\ref{eq:energetic}) and~\eqref{eq:fast} can of course be imposed on the initial distribution of particle momenta, given the distribution $\hat\rho$ of scatterer energies. Provided that, in the distribution to which the particle eventually equilibrates, almost all of the particles continue to satisfy these conditions, the
following analysis will provide an accurate description of the equilibration process. Since, as we will show, the particle distribution equilibrates to a final temperature $T$ for which $\|p\|^2$ is of the order $k_{\mathrm{B}}T\sim \overline{P^2}\sim E_*$  [see~\eqref{eq:betastar}], both of these conditions do indeed continue to be satisfied during the particle's approach to equilibrium.

Since we will focus on fast particles, we expand the function $R\left( p,\kappa\right) $ in
inverse powers of $\Vert p\Vert$. This leads to a corresponding expansion
\begin{equation}
\Delta E\left( p,\kappa\right) =\sum_{\ell=0}^{K}\frac{\beta^{\left(
\ell\right) }\left( \kappa\right) }{\Vert p\Vert^{\ell}}+\mathrm{O}\left(
\Vert p\Vert^{-K-1}\right) ,  \label{eq:deltaEexpansion}
\end{equation}
for the energy transferred in a single collision, in which the $%
\beta^{\left( \ell\right) }\left( \kappa\right) $ are explicit scalar
functions of the incoming collision parameters. We first remark that, since $t_+$ is
of order $\|p\|^{-1}$, it is clear that $\beta^{(0)}=0$. A straightforward computation of $\beta^{(1)}$ and $\beta^{(2)}$, worked out in the Appendix, leads to the following results:
\begin{equation}\label{eq:beta1and2}
\left.
\begin{split}
\beta^{(1)}(\kappa)=&\alpha \eta'(Q)\frac{P}{M}L_0(\|b\|)\\
\beta^{(2)}(\kappa)=&\frac{\alpha }{M^2} \left[P^2\eta''(Q)-M\eta'(Q)U'(Q)\right]L_1(\|b\|)\\
&\qquad\qquad\qquad\qquad\qquad-\frac{\alpha\tilde\alpha}{2M}(\eta'(Q))^2L_{0}^{2}(\|b\|),
\end{split}
\right\}
\end{equation}
where for $k\in \mathbb{N}$, and $b,e\in \mathbb{R}^{d}$, with $e\cdot e=1$,
$b\cdot e=0$,
\begin{equation}\label{eq:Lk}
L_{k}(\|b\| ):=\int_{0}^{1}\mathrm{d}\lambda \ \lambda ^{k}\sigma
(b+(\lambda -\frac{1}{2})e).
\end{equation}
Note that the right hand side in~\eqref{eq:Lk} is a function of $\Vert b\Vert $ only, as a result of the rotational
invariance of $\sigma $.
It follows that [see \eqref{eq:rhocond} and \eqref{eq:overlinedef}]
\begin{equation}\label{eq:beta1averages}
 \overline{\beta^{(1)}}=0, \qquad
\Sigma _{1}^{2}:=\overline{\left( \beta ^{(1)}\right) ^{2}}
=\frac{\alpha^2}{M^2}\ \overline{(\eta'P)^2}\ \overline{L_0^2},
\end{equation}
since $\rho(Q,P) = \rho(Q,-P)$. Also, since we assumed that $\rho$ is stationary for
the free dynamics of the scatterer generated
by $H_{\mathrm{scat}}$ [see~(\ref{eq:rhocond})], one readily checks that,
for any choice of $\eta$,
\begin{equation*}
\overline{\left[P^2\eta''(Q)-M\eta'(Q)U'(Q)\right]}=0.
\end{equation*}
Hence
\begin{equation}\label{eq:beta2average}
  \overline{\beta^{(2)}}=-\frac{\tilde\alpha\alpha}{2M}\ \overline{(\eta' L_0)^2}.
\end{equation}

Assuming $\|p_n\|$ is large in the sense of~\eqref{eq:energetic}-\eqref{eq:fast}, and dropping for the moment all higher order terms, we conclude from~\eqref{eq:deltaEexpansion} that the energy change undergone by the particle during the $n$th scattering event is approximately given by:
\begin{equation}\label{eq:deltaEndominant}
\Delta E_n:=\Delta E\left( p_n,\kappa_n\right)=\frac{\beta^{\left(
1\right) }\left( \kappa_n\right) }{\Vert p_n\Vert}+\frac{\beta^{\left(2\right) }\left( \kappa_n\right) }{\Vert p_n\Vert^{2}}.
\end{equation}
One recognizes here a dominant fluctuating term (in $\|p_n\|^{-1}$), independent of the value of $\tilde\alpha$, which is of zero average in view of~\eqref{eq:beta1averages}, and a smaller subdominant term (in $\|p_n\|^{-2}$) which fluctuates about a \emph{negative} non-zero average when $\tilde\alpha=\alpha$.  Whereas the random fluctuations of the first term have a tendency to increase the particle's speed without bound, the second term, while weaker, is systematic and has a tendency to reduce it. It is the competition between these two effects when $\tilde\alpha=\alpha$ that eventually leads the particle to equilibrate, as we show below. If, on the other hand, $\tilde\alpha=0$, then, as shown in~\cite{adblafp}, the particle undergoes a stochastic acceleration, with it's speed
increasing as $\|p_n\|\sim n^{1/3}$.

Note, therefore,  that for this class of systems, we have been able to explicitly separate the force acting on the particle as the result of its passage through the medium into a frictional and a random part, a notoriously difficult problem of statistical mechanics in general (see~\cite{kuboII}, p.37). It is furthermore clear from the above discussion that the frictional part is due to a back-reaction effect: it results from the change induced in the particle's motion by the change in the medium's motion, which is itself brought about by the passage of the particle. This effect is entirely absent when $\tilde\alpha=0$.

We now prove that the two competing effects described above balance out so as to drive the particle's momentum distribution precisely to a Maxwell-Boltzmann distribution. It is convenient for analyzing the asymptotic behavior to focus on the random walk associated with a new scaled dynamical variable
\begin{equation}
\xi_n=\frac{\Vert p_{n}\Vert ^{3}}{3\Sigma _{1}}.  \label{eq:xidef}
\end{equation}
For that purpose, we first note, using (\ref{eq:deltaEexpansion}),  that
\begin{equation*}
\frac{\Vert p_{n+1}\Vert ^{2}}{\Vert p_{n}\Vert ^{2}}=1+\sum_{\ell=1}^{4}\frac{%
2\beta _{n}^{\left( \ell\right) }}{\Vert p_{n}\Vert ^{\ell+2}}+\mathrm{O}\left(
\Vert p_{n}\Vert ^{-7}\right) ,
\end{equation*}%
where $\beta _{n}^{(\ell )}:=\beta ^{(\ell)}(\kappa _{n}).$ From this last
expression we obtain the relations%
\begin{align*}
\frac{\Vert p_{n+1}\Vert }{\Vert p_{n}\Vert }& =1+\sum_{\ell=1}^{3}\frac{\beta
_{n}^{\left(\ell\right) }}{\Vert p_{n}\Vert ^{\ell+2}}+\mathrm{O}\left( \Vert
p_{n}\Vert ^{-6}\right) \\
\Vert p_{n+1}\Vert -\Vert p_{n}\Vert & =\sum_{\ell=1}^{3}\frac{\beta
_{n}^{\left(\ell\right) }}{\Vert p_{n}\Vert ^{\ell+1}}+\frac{\beta _{n}^{\left(
4\right) }-\frac{1}{2}\left( \beta _{n}^{\left( 1\right) }\right) ^{2}}{%
\Vert p_{n}\Vert ^{5}}+\mathrm{O}\left( \Vert p_{n}\Vert ^{-6}\right) ,
\end{align*}%
and consequently
\begin{align}
\Delta \Vert p_{n}\Vert ^{3}& =\Vert p_{n}\Vert ^{2}\Delta \Vert p_{n}\Vert
\left[ 1+\frac{\Vert p_{n+1}\Vert }{\Vert p_{n}\Vert }+\frac{\Vert
p_{n+1}\Vert ^{2}}{\Vert p_{n}\Vert ^{2}}\right]  \notag \\
& =\sum_{\ell=1}^{3}\frac{3\beta _{n}^{\left( \ell\right) }}{\Vert p_{n}\Vert
^{\ell-1}}+\frac{3\left( \beta _{n}^{\left( 4\right) }+\frac{1}{2}\left( \beta
_{n}^{\left( 1\right) }\right) ^{2}\right) }{\Vert p_{n}\Vert ^{3}}+\mathrm{O%
}\left( \Vert p_{n}\Vert ^{-4}\right) .  \label{eq:deltapcubed}
\end{align}%
In terms of $\xi_n$ defined in~\eqref{eq:xidef} and the parameters%
\begin{align*}
\epsilon _{n}& =\frac{\beta _{n}^{\left( 1\right) }}{\Sigma _{1}},\qquad \mu
_{n}=-(3\Sigma _{1}^4)^{-1/3}\beta _{n}^{(2)}, \\
\nu _{n}& =3^{-2/3}\Sigma _{1}^{-5/3}\beta _{n}^{(3)},\qquad \gamma _{n}=%
\frac{1}{3}\left( \frac{\beta _{n}^{(4)}}{\Sigma _{1}^{2}}+\frac{1}{2}%
\epsilon _{n}^{2}\right) ,
\end{align*}%
equation~(\ref{eq:deltapcubed}) takes the form%
\begin{equation}
\xi _{n+1}=\xi _{n}+\epsilon _{n}-\mu _{n}\xi _{n}^{-1/3}+\nu _{n}\xi
_{n}^{-2/3}+\gamma _{n}\xi _{n}^{-1}
+\mathrm{O}(\xi _{n}^{-4/3}). \label{eq:xirw0}
\end{equation}%
From~\eqref{eq:beta1averages}-\eqref{eq:beta2average} one finds
\begin{equation}\label{eq:epsilonmuaverages}
\overline{\epsilon _{n}} =0,\qquad \overline{\epsilon _{n}^{2}}=1,\quad
\overline{\mu _{n}} :=\overline{\mu }=\frac{%
\tilde{\alpha}\alpha }{2(3\Sigma _{1}^4)^{1/3}M}\overline{(\eta ^{\prime }L_{0})^{2}}\geq0.
\end{equation}
The rather involved computation of the higher order coefficients $\beta^{(3)}_n$ and $\beta^{(4)}_n$ is performed in the
Appendix and leads to the result that [see~\eqref{eq:beta3av} and~\eqref{eq:overlinebeta4}]
\begin{equation*}
\overline{\nu _{n}} =0,\
\end{equation*}%
and that
\begin{equation}\label{eq:overlinegamma}
\overline{\gamma _{n}}:=\overline{\gamma }=\frac{1}{3}\left( \frac{\overline{%
\beta _{n}^{(4)}}}{\Sigma _{1}^{2}}+\frac{1}{2}\right) =\frac{1}{6}\left(
d-2\right) +\frac{\overline{\delta \beta _{a}^{(4)}+\delta \beta _{b}^{(4)}}%
}{3\Sigma _{1}^{2}},
\end{equation}%
where $\delta \beta _{a}^{(4)}$ and $\delta \beta _{b}^{(4)}$ are defined
in~(\ref{eq:deltabetaa4}) and~(\ref{eq:deltabetab4}) of the Appendix.
Using these results (\ref{eq:xirw0}) can be written%
\begin{equation}
\xi _{n+1}=\xi _{n}+\epsilon _{n}-\overline{\mu }\xi _{n}^{-1/3}+\overline{%
\gamma }\xi _{n}^{-1}+\mathrm{O}_{0}(\xi _{n}^{-1/3})+\mathrm{O}(\xi
_{n}^{-4/3}),  \label{eq:xirw1}
\end{equation}%
where the notation $\mathrm{O}_{0}\left( \Vert \xi _{n}\Vert ^{-1/3}\right) $
means the term is $\mathrm{O}\left( \Vert \xi _{n}\Vert ^{-1/3}\right) $ and
of zero average. Again focusing on particles that over the course of their
evolution spend an overwhelming amount of time at high speeds, we drop the
asymptotically small error terms in this last equation to obtain a
one-dimensional random walk%
\begin{equation}
\xi _{n+1}=\xi _{n}+\epsilon _{n}-\frac{\overline{\mu }}{\xi _{n}^{1/3}}+%
\frac{\overline{\gamma }}{\xi _{n}}.  \label{eq:xirw}
\end{equation}%
Comparing this to the random walk for $\|p_n\|^2$ in~\eqref{eq:deltaEndominant}, one recognizes again in the second and third term on the right hand side the effect of the dominant fluctuating part of the force and of its subdominant frictional part. The fact that, in~\eqref{eq:xirw}, the dominant fluctuating term is independent of the random variable $\xi_n$ itself constitutes an advantage over~\eqref{eq:deltaEndominant} that will simplify the following analysis. It arises from the fact that $\xi_n$
is proportional to the third power of the particle's speed.  Again, we stress the all-important strict {positivity} of $\overline{\mu}$  [see~\eqref{eq:epsilonmuaverages}] that occurs when $\tilde\alpha=\alpha$ and that, as we will see, is needed to assure that the third term on the right hand
side of (\ref{eq:xirw}) acts as a source of dynamical friction capable of balancing the
diffusive growth of $\xi_n$ generated by the independent and identically distributed terms $\epsilon_n$ of zero mean.
The crucial role of the much smaller last term in~(\ref{eq:xirw}) will become evident
below. We note that essentially the same equation was obtained in~\cite{adblafp} for the non-reactive case for which $\tilde\alpha=0$, and so $\overline\mu=0$. It was then proven for that case that $\xi_n\sim\sqrt{n}$, leading to unbounded growth of the energy of the particle. This stochastic acceleration is completely suppressed when $\overline\mu>0$, as we will see.

The value of $\overline{\gamma }$ in~\eqref{eq:overlinegamma} is obtained as the result of an
involved computation of the coefficient $\overline{\beta ^{(4)}}$ in the
Appendix. In particular it is shown there that, up to corrections that are small provided $\alpha$ is small and that the bath is sufficiently energetic (i.e., that $E_*$ is sufficiently high), $\overline{\gamma }$ depends on the dimensionality $d$ of the
system through the simple relation
\begin{equation}
\overline{\gamma }=\frac{1}{6}(d-2),  \label{eq:overlinegammab}
\end{equation}%
which is independent of the model parameters, the interaction $\eta $, and the
potential $U$.

We now study the asymptotic behavior of the random walk (\ref{eq:xirw})
executed by the variable $\xi_n$ for the case in which $\overline\mu>0$. For
fast particles
relevant to the
present analysis, and small coupling $\alpha$, the dynamical
quantity $\xi_n\propto$ $\Vert p_{n}\Vert^{3}/\alpha^2$
will be much greater than unity, and the changes $\Delta\xi_n$ that occur
during any given collision will be small compared to the value of $\xi_n$
itself. Thus, over a large number of collisions during which $\xi_n$ does not
appreciably change, we may take $n$ to be a quasi-continuous variable, in
terms of which the random walk (\ref{eq:xirw}) takes the form of a
stochastic differential equation
\begin{equation}
\dot{\xi}(n)=\epsilon(n)-v^{\prime}\left( \xi(n)\right) ,  \label{eq:langevin}
\end{equation}
where $\epsilon(n)$ is the random process satisfying
\begin{equation*}
\langle \epsilon(n)\rangle=0,\quad\langle \epsilon(n)\epsilon(n^{\prime})\rangle=\delta
(n-n^{\prime}),
\end{equation*}
and where
\begin{equation}
v(\xi)=\frac{3}{2}\overline{\mu}\ \xi^{2/3}-\overline{\gamma}\ln \xi.
\label{eq:v}
\end{equation}
The corresponding forward Kolmogorov equation for the
momentum density $\tilde{\rho}(\xi,n)$ of the particle then reads
\begin{equation}
\partial_{n}\tilde{\rho}(\xi,n)=\partial_{\xi}\left[ v^{\prime}(\xi)\tilde{\rho}(\xi,n)+\frac{1}{2}%
\partial_{\xi}\tilde{\rho}(\xi,n)\right] .  \label{eq:FP}
\end{equation}
We point out that (\ref{eq:langevin}) can alternatively be interpreted as
a Langevin-type equation describing the rate of change of the position $\xi$ of an overdamped particle
moving in a confining potential $v$ and subject to a fluctuating random
force $\epsilon(n)$. In that interpretation the variable $n$ in (\ref{eq:langevin})
plays the role of time, and~\eqref{eq:FP} is the Fokker-Planck equation for the probability
distribution $\tilde\rho(\xi,n)$ associated with such a Langevin process.
It is then clear that the solutions of~\eqref{eq:FP} converge for large $n$ (i.e., after many
collisions) to the limiting distribution
\begin{align}
\tilde{\rho}_{\mathrm{eq}}(\xi) &
=\tilde{\rho}_{\mathrm{eq}}\xi_{0})\exp\left[ -2\int_{\xi_{0}}^{\xi}\mathrm{d}\xi^{\prime
}v^{\prime}(\xi^{\prime})\right] =\mathcal{N}\exp\left[ -2v(\xi)\right]
\notag  \label{eq:stationarysol} \\
& =\mathcal{N\;}\xi^{2\overline{\gamma}}\exp(-3\overline{\mu}\ \xi^{2/3}),
\end{align}
where $\mathcal{N}$ is a normalization constant.
Physically, $\tilde{\rho}_{\mathrm{eq}%
}(\xi)d\xi$ gives, in equilibrium, the fraction of collisions
for which the final particle speed $\Vert p\Vert$
is such that the quantity $\xi=\frac{1}{3}\Vert p_{n}\Vert^{3}/\Sigma_{1}$ has
a value lying between $\xi$ and $\xi+d\xi.$ Changing variable from $\xi$ back to the
speed $\Vert p\Vert$, one has, in general, $\tilde{\rho}(\Vert p\Vert,n)=\Vert p\Vert
^{2}\Sigma_{1}^{-1}\tilde{\rho}(\xi,n)$, which gives the distribution%
\begin{equation}
\tilde{\rho}_{\mathrm{eq}}(\Vert p\Vert)=\frac{\Vert p\Vert^{2}}{%
\Sigma_{1}}\tilde{\rho}_{\mathrm{eq}}(\xi)  \label{eq:fromntot}
\end{equation}
associated with the fraction of collisions, in equilibrium, in which the final particle
speed is $\Vert p\Vert$. The average \emph{time} that a particle
emerging from a collision with speed $\Vert p\Vert$ remains at that speed after
the collision is just
$\ell_{\ast}/\Vert p\Vert$. Thus, the
equilibrium probability density $\rho_{\mathrm{eq}}(\Vert p\Vert)$,
which governs the fraction
$\rho_{\mathrm{eq}}(\Vert p\Vert)dp$ of \emph{time}
that a particle spends in equilibrium with a speed lying
between $\Vert p\Vert$ and $\Vert p\Vert+d\Vert p\Vert$,
is found from the relation
\begin{equation*}
\rho_{\mathrm{eq}}(\Vert p\Vert)\propto\frac{\ell_{\ast}}{\Vert p\Vert }\tilde{\rho}_{%
\mathrm{eq}}(\Vert p\Vert).
\end{equation*}
Using (\ref{eq:overlinegammab}), we thus find that the
distribution $\rho_{\mathrm{eq}}(\Vert p\Vert,t)$ of particle speeds $\Vert
p\Vert$ asymptotically approaches a Maxwellian
\begin{eqnarray}
\rho_{\mathrm{eq}}(\Vert p\Vert)&=&\mathcal{N}_{0}\Vert_{{}}p\Vert^{1+6\overline {%
\gamma}}\exp(-\frac{\Vert p\Vert^{2}}{2k_{\mathrm B}T})\nonumber\\
&=&\mathcal{N}_{0}\;\Vert
p\Vert^{d-1}\exp(-\frac{\Vert p\Vert^{2}}{2k_{\mathrm B}T}),  \label{eq:Peq}
\end{eqnarray}
with an effective  temperature $T$ given
by the expression
\begin{equation}\label{eq:betastar}
\frac12 k_{\mathrm{B}}T=-\frac{(3\Sigma_{1})^{2/3}}{6\overline{\mu}}=\frac{\overline{(\beta^{(1)})^{2}}}{
\left[-\overline{\beta^{(2)}}\right] }=\frac{1}{2M}\frac{\overline{{\eta^{\prime}}^{2}P^{2}}}{\overline{{\eta^{\prime}}^{2}}},
\end{equation}
the meaning of which we shall further analyze in Sec.~\ref{s:numerics}. Note the crucial role played by the precise value of $\overline{\gamma}=\frac16(d-2)$ appearing in the last term of~\eqref{eq:xirw}; it produces the correct ``density of states''
prefactor in~(\ref{eq:Peq}), without which the
particle speed distribution $P_{\mathrm{eq}}(\Vert p\Vert)$ would not
strictly be Maxwellian. To finally show that the
distribution of (vector) momenta $p$
approaches the corresponding Maxwellian distribution
\begin{equation}\label{eq:maxwell}
\rho_{\mathrm{eq}}(p)=\frac{1}{\left(2\pi k_{\mathrm B}T\right)^{\frac{d}{2}}}\exp(-\Vert p\Vert^{2}/2k_{\mathrm B}T),
\end{equation}
it is sufficient to prove that moving particles, as a result of
repeated collisions, asymptotically lose any memory of their initial
direction, so that the corresponding momentum distribution becomes
asymptotically isotropic. Such a demonstration is given in Sec.~\ref%
{s:diffusion}, which contains an analysis of the rate at which random
collision-induced deflections turn the particle, and in which that
information is then further used to calculate the diffusion constant that characterizes
the growth of the particle's mean-squared displacement in equilibrium. Before turning to an analysis of the spatial motion, however, we consider in Sec.~\ref{s:numerics} some specific consequences of our analysis of the particle's approach to
equilibrium, and illustrate some of the predictions of our analysis with
appropriate numerical calculations.


\section{Generalized equipartition and numerical simulations}\label{s:numerics}

We begin by noting that the expression for the effective temperature $T$ in~(\ref{eq:betastar}) gives rise to a generalized equipartition relation: for $i=1\dots d$
\begin{equation}\label{eq:Tweighteddav}
\frac12{\left<{p_i^2}\right>_{T}}=\frac12 k_{\mathrm{B}}T=\left\langle \frac{P^{2}}{2M}\right\rangle _{\eta}:=%
\frac{1}{2M}\;\frac{\;\overline{\eta^{\prime}{}^{2}P^{2}}\;}{\overline {%
\eta^{\prime}{}^{2}}}.
\end{equation}
Here $\left<\cdot\right>_{T}$ stands for an average with respect to
the Maxwell distribution~\eqref{eq:maxwell} and $\left<\cdot\right>_\eta$ denotes
a  \emph{weighted} average of the kinetic energy of the scatterers in the medium, calculated using the
weighting function $[\eta'(Q)]^2 \rho\left( Q,P\right) ,$ rather than
simply using the stationary scatterer distribution $\rho\left( Q,P\right)$
alone.
Clearly, in the
case in which the coupling is linear, meaning that
$\eta(Q)=Q$, \eqref{eq:Tweighteddav} reduces to the standard equipartition result
\begin{equation}
\frac12{\left<{p_i^2}\right>_{T}}=\frac{1}{2}kT=\frac{\overline{P^{2}}}{2M},  \label{eq:naturaltemp}
\end{equation}
according to which all degrees of freedom of the system have the same mean kinetic energy.
Note that this result for linear coupling is completely
independent of the stationary scatterer distribution $\rho:$ the medium does
not itself need to be in thermal equilibrium for (\ref{eq:naturaltemp}) to
obtain.

If the coupling is non-linear, so that $\eta^{\prime}\not =1$, the situation
is more complicated, as we have indicated. Nevertheless, for the special
case in which the scatterers are themselves in thermal equilibrium at some
temperature $T_*$, the corresponding canonical distribution $Z_{
\mathrm{scat}}^{-1}\exp(-H_{\mathrm{scat}}(Q,P)/k_{\mathrm B}T_*)$ factors into
separate distributions for $Q$ and $P$, and hence, in this case, so does the
average
\begin{equation*}
\overline{\eta^{\prime}{}^{2}P^{2}}=\overline{\eta^{\prime}{}^{2}}\;
\overline{P^{2}}.
\end{equation*}
Thus, in this way the dependence on $\eta^{\prime}$ disappears, and one
again finds that the particle momentum distribution converges to a
Maxwellian with a temperature $T=T_*$ given by~(\ref{eq:naturaltemp}), independent of the precise form of the coupling.
As a result, standard equipartition holds, meaning that asymptotically in time all degrees of
freedom of the system have the same average kinetic energy, given by $\frac12 k_{\mathrm B}T$.

\begin{figure}[t]
\hspace{-1cm}
\includegraphics[height=6.75cm,keepaspectratio]{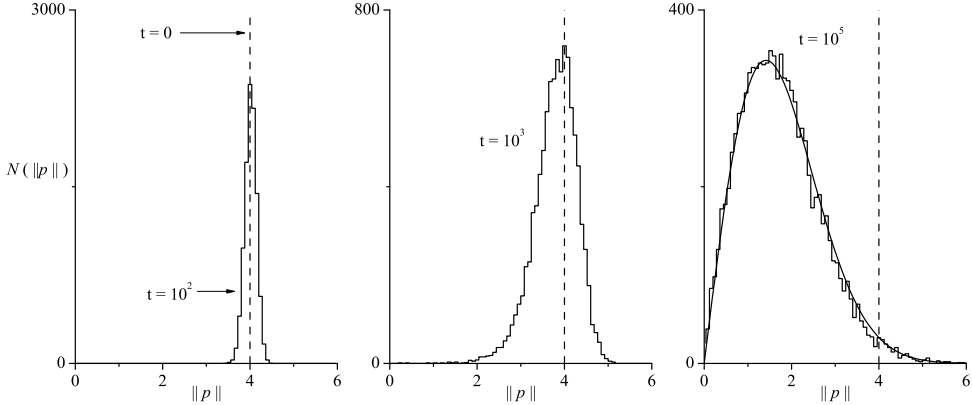}
\caption{Approach to equilibrium of the particle speed distribution for a
particle moving through a lattice of thermally equilibrated harmonic
oscillators with average energy $E_*=2$, with linear coupling. The solid
curve in the last panel is
a Maxwellian velocity distribution with the predicted final
thermal energy $kT = 2$, equal to that of the oscillators.}
\label{fig:approachCLIN}
\end{figure}

\begin{figure}[t]
\hspace{-1cm}
\includegraphics[height=6.95cm,keepaspectratio=true]{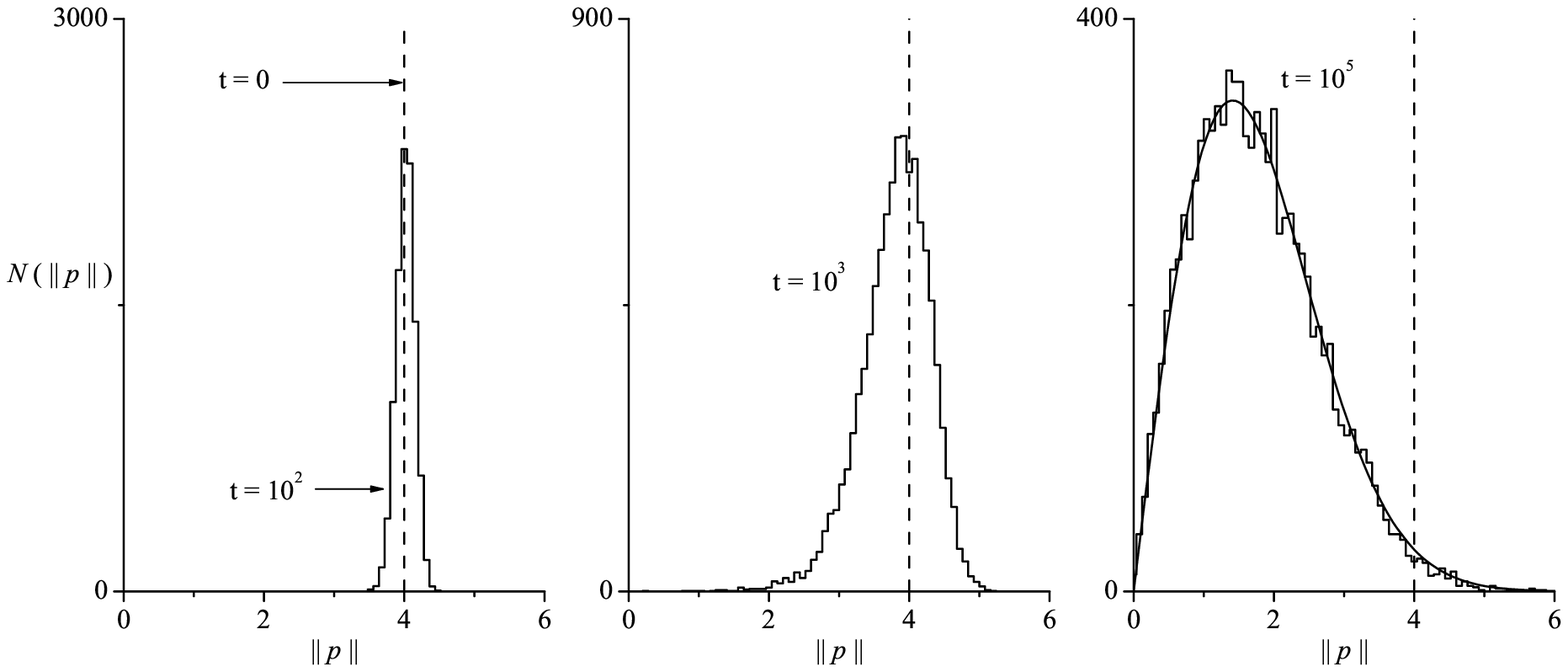}
\caption{Approach to equilibrium of the particle speed distribution for a
particle moving through a lattice of
harmonic oscillators each drawn from their own microcanonical distribution
with average energy $E_*=2$, and
linear coupling. The solid curve
in the last panel is
a Maxwellian distribution with the predicted final thermal energy $kT = 2$.}
\label{fig:approachMLIN}
\end{figure}

As an interesting example in which the effective temperature $T$ that
is reached does \emph{not} lead to the standard equipartition relation,
consider a situation in which the scatterers are oscillators with $U(Q)=
\frac{1}{2}Q^{2}$, each drawn from a uniform ensemble with phase space density $
\rho(Q,P)\sim\delta (H_{\mathrm{scat}}(Q,P)-E_*)$ of
fixed energy $E_*,$ and quadratically coupled
to the particle with $\eta(Q)=\frac{1}{2}Q^{2}$. A simple computation shows
that for this case, for all $i=1\dots  d$,
\begin{equation}
\frac12{\left<{p_i^2}\right>_T}=\frac{1}{2}kT=\frac{1}{2M}\frac{\;\;\overline{Q^{2}P^{2}}\;\;}{%
\overline{Q^{2}}}=\frac{E_*}{4}=\frac{1}{2}\frac {\overline{P^{2}}
}{2M}.  \label{eq:quadtemp}
\end{equation}
In other words, the mean kinetic energy per degree of freedom of the particle equals one half
the mean kinetic energy of the scatterer: standard equipartition does therefore not hold in this case.

To demonstrate the general features that emerge from the preceding analysis,
we have performed numerical calculations for particle-scatterer systems
evolving according to the actual equations of motion (\ref{eq:eqmotion}),
for the two cases of linear coupling $\eta (Q)=Q$ and quadratic coupling $%
\eta (Q)=\frac{1}{2}Q^{2}$ discussed above. All numerical calculations were
performed with particles moving through a two-dimensional hexagonal array of
harmonic scatterers $(U=\frac{1}{2}Q^{2}),$ centered at the points $%
x_{N}=N_{1}u+N_{2}v,$ where $u=(a_{\ast },0),\ v=(a_{\ast }/2,\sqrt{3}%
a_{\ast }/2)$, and $N=\left( N_{1},N_{2}\right) \in \mathbb{Z}^{2}.$ For
these calculations the form factor was chosen to be $\sigma (q)=\chi \left(
1-2\Vert q\Vert \right) $ where $\chi (x)$ is the usual step function, equal
to unity for $x> 0$ and to zero for $x\leq 0$, and the lattice parameter $%
a_{\ast }$ was set equal to $10/9,$ ensuring a finite horizon.  Note that with
this choice of a form factor, as can be seen in the close-up in
Fig.~\ref{fig:typicaltraj}, the particle undergoes (easily computed)
impulsive changes in its momentum at the instants it encounters
the edge of an interaction
region, but at all other times follows straight line trajectories at
fixed speed.
\begin{figure}[t]
\hspace{-1cm}
\includegraphics[height=7cm,keepaspectratio=true]{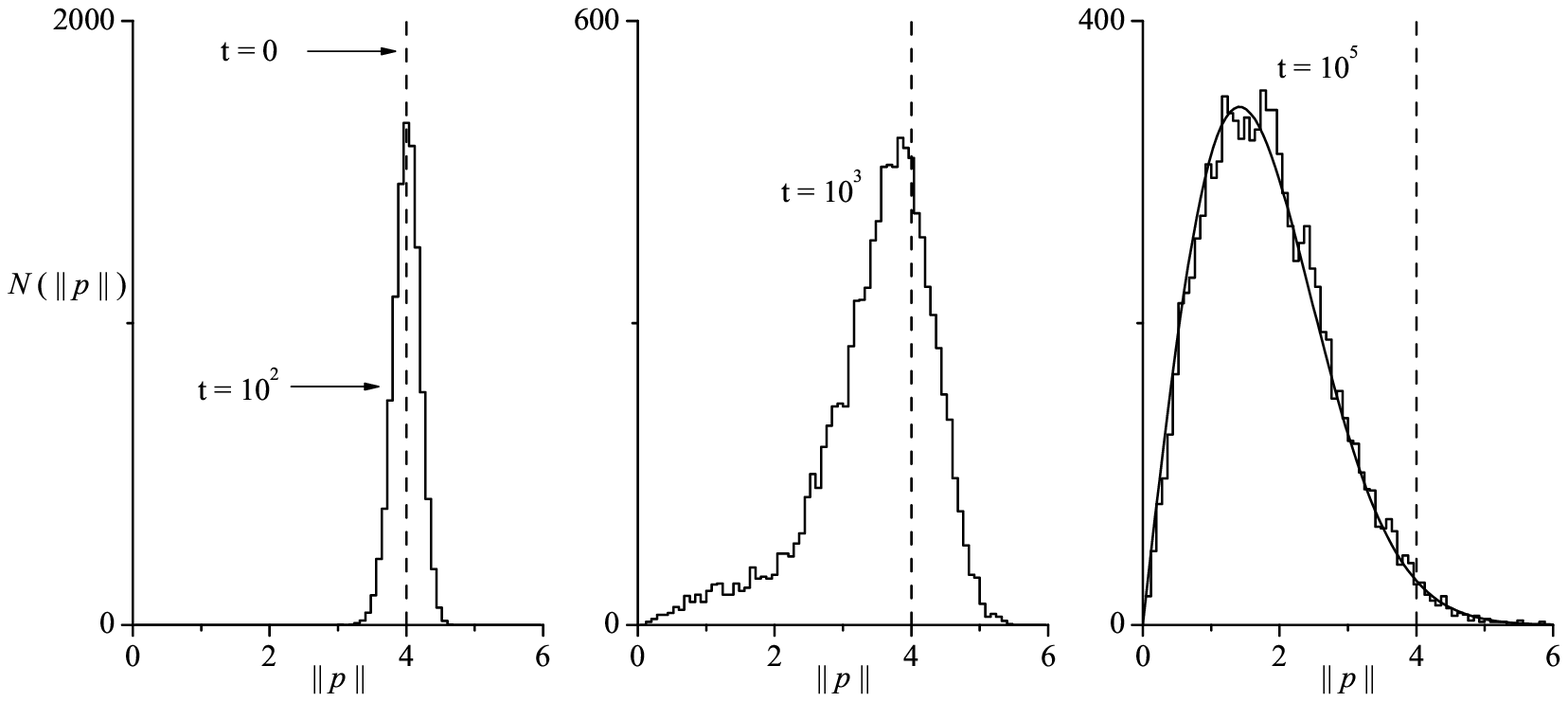}
\caption{Approach to equilibrium of the particle speed distribution for a
particle moving through a lattice of thermally distributed
harmonic oscillators with average energy $E_*=2$, with quadratic coupling.
The solid curve in the last panel is
a Maxwellian velocity distribution with a thermal energy $kT = 2$, equal to that
of the oscillators.}
\label{fig:approachCQUAD}
\end{figure}

\begin{figure}[t]
\hspace{-1cm}
\includegraphics[height=7cm,keepaspectratio=true]{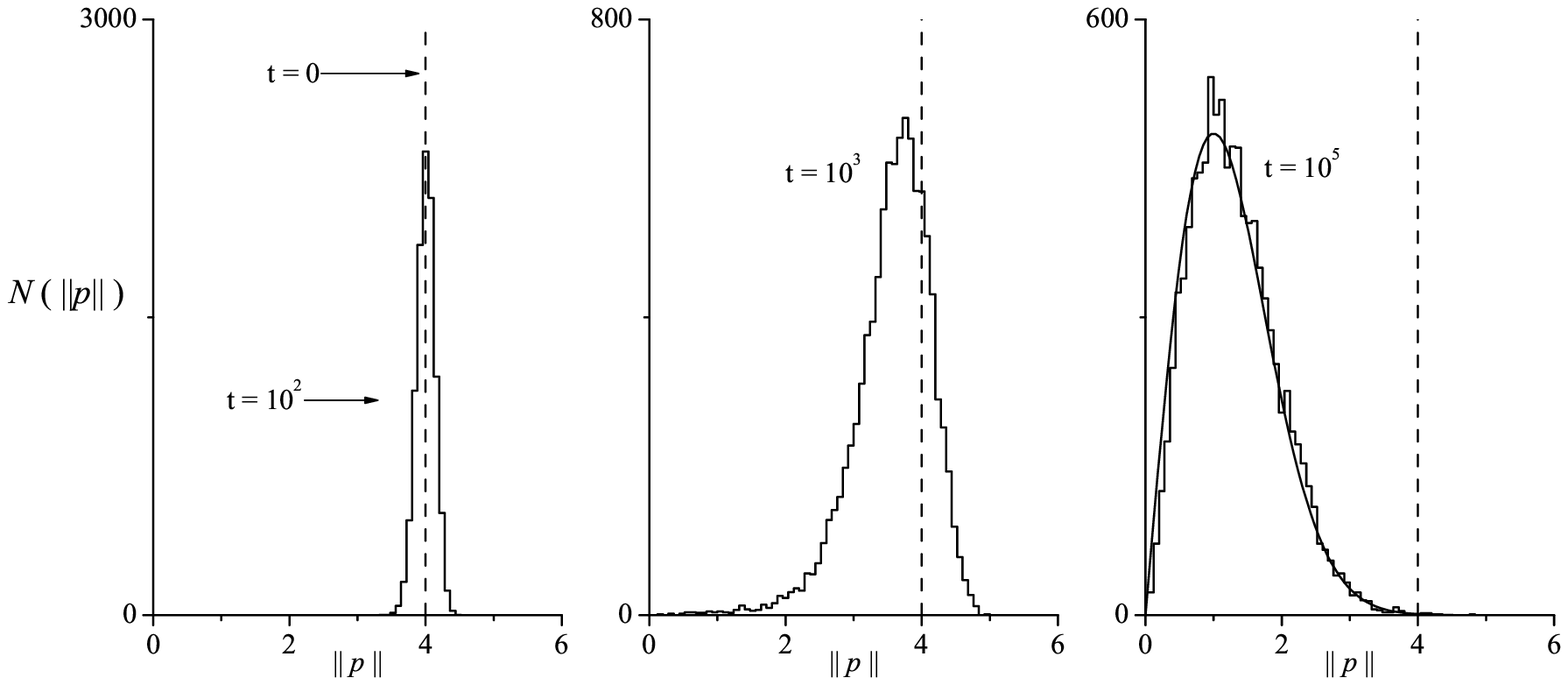}
\caption{Approach to equilibrium of the particle speed distribution for a
particle moving through a lattice of
harmonic oscillators each drawn from their own microcanonical distribution
with average energy $E_*=2$, and
quadratic coupling. The
solid curve in the last panel is
a Maxwellian velocity distribution with a thermal energy $kT = 1$ predicted by
analysis, one-half the value found in Figs. \ref{fig:approachCLIN}-\ref{fig:approachCQUAD}.}
\label{fig:approachMQUAD}
\end{figure}

In the
numerical results presented
in Figs.~\ref{fig:approachCLIN}-\ref{fig:approachMQUAD},
for a given set of parameters, each
particle in an ensemble of $10^{4}$ trajectories was given the same initial
speed, at a random point on the boundary of the scatterer at the origin,
with an initial velocity drawn with equal probability from all physically
possible outward directions. A typical time sequence showing approach to
equilibrium is presented in Fig.~\ref{fig:approachCLIN} for a
system with
linear coupling, $\alpha =0.1,$ and oscillators of unit mass ($M=1$)
initially in thermal equilibrium at a temperature such that $E_*=kT=2.$ The
histograms in
this figure show the evolution of an initially sharp distribution of
particle speeds (indicated by the dashed vertical line) into the predicted 2-%
$d$ Maxwell speed distribution (shown as a smooth solid curve in the last
panel of that figure) with
a final effective temperature $T_{\ast }=T=2,$ as predicted by the analysis
given above.

A similar sequence is depicted in Fig.~\ref{fig:approachMLIN} for the same system, but
with
the initial state of each oscillator drawn from a uniform distribution on the energy surface
$H_{\mathrm{scat}}(Q,P)=E_*=2$.
The last panel of that figure
confirms the prediction
that for linear coupling the final distribution
is
the same as when the oscillators
are thermally distributed. Additional
numerical results
(not shown here)
confirm that this
limiting distribution is independent
of the strength of the
coupling parameter $\alpha $
as predicted by~\eqref{eq:betastar}. Figures~\ref{fig:approachCQUAD}
and \ref{fig:approachMQUAD} show corresponding results
for the case of quadratic coupling [$\eta (Q)=Q^{2}/2$],
again with $\alpha =0.1$, $M=1$, and $E_*=2$.
With quadratic coupling, the preceding
analysis predicts different final temperatures according to whether the oscillators
are distributed thermally
with average energy~$E_*$
or uniformly on their individual energy surfaces $H_{\mathrm{scat}}(Q,P)=E_*=2$.
The numerical results confirm
this prediction. The limiting temperature of the particles that is
obtained when the scatterers
are uniformly distributed, given in~(\ref{eq:quadtemp}), is one-half the
value that emerges when the scatterers are drawn from a
canonical distribution having the same average kinetic energy.
The standard equipartition relation does not hold, in this case, since
the average kinetic energy associated with the particle's degrees of freedom
equals one half that associated with the oscillator.


\section{Diffusion and suppression of stochastic acceleration}\label{s:diffusion}
We now study the motion in position space of an ensemble of particles in
thermal equilibrium at high temperature, each of which undergoes the random walk~\eqref{eq:finalrw} under the
same hypotheses on the model as in the previous section, in particular that $\tilde\alpha=\alpha$. Our
analysis closely follows the approach of~\cite{adblafp}, where the case $\tilde\alpha=0$ was treated.
At the end of the section we compare the behavior of the mean-squared displacement in these
two very different situations.

We begin by noting that, since $y_{n+1}=y_n+\ell_*e_n$ and $e_n=\frac{p_n}{\|p_n\|}$, an understanding
of the particle's trajectory requires more than just a knowledge of how its speed changes in time;
one also needs to understand how the random changes induced in the particle's momentum by the scatterers
cause it to turn.  To explore this question, we use perturbation
theory in~(\ref{eq:onehitequation}), to expand the function
\begin{equation}  \label{eq:Rexpansion}
R\left( p, \kappa\right) =\sum_{k=1}^{K} \frac {\alpha^{\left( k\right)
}\left( e, \kappa\right) }{\|p\|^{k}} +\mathrm{O}\left( \|p\|^{-K-1}\right)
,\quad e=\frac{p}{\|p\|}.
\end{equation}
in inverse
powers of $\|p\|$.
In particular, one finds that
\begin{equation}  \label{eq:alpha1}
\alpha^{\left( 1\right) }\left( e, \kappa\right)
=-\alpha\eta(Q)\int_{-\infty}^{+\infty} \mathrm{d} \lambda\ \nabla
\sigma(b+\lambda e), \ \mathrm{so\ that}\ e\cdot\alpha^{(1)}=0.
\end{equation}
Starting from the first equation of (\ref{eq:finalrw}) and (\ref%
{eq:Rexpansion}), a simple computation \cite{adblafp} yields the result
\begin{equation}  \label{eq:enplus1}
e_{n+1}=e_{n}+\delta_{n},
\end{equation}
where
\begin{equation*}
\delta_{n}=\frac{\alpha^{\left( 1\right) }_{n}}{\|p_{n}\|^{2}} + \mathrm{O}%
(\|p_{n}\|^{-3}),\quad\alpha_{n}^{(1)}:=\alpha^{(1)}(e_{n},\kappa_{n}).
\end{equation*}
Thus, as suggested in Sec. \ref{s:equilibration}, an initially fast particle of
momentum $p_{0}$ undergoes small random and
independent deflections that cause the unit vectors $e_{n}$ to diffuse
isotropically over the
unit sphere. To determine the number of collisions required for the
diffusing unit vectors to cover the sphere, we compute the
value $m=m_{*}(p_0)$ for which the correlation function
\begin{equation*}
\left\langle \|e_{m}-e_{0}\|^{2}\right\rangle =\sum_{k=0}^{m-1}\sum
_{k^{\prime}=0}^{m-1}\left\langle
\delta_{k}\cdot\delta_{k^{\prime}}\right\rangle
\end{equation*}
becomes of order $1$. Under the assumptions of
the random walk~\eqref{eq:finalrw}, the
off-diagonal terms of this last expression clearly have zero average so that
\begin{equation*}
\left\langle \|e_{m}-e_{0}\|^{2}\right\rangle =m\frac{\overline{
\|\alpha_{0}^{\left( 1\right) }\|^{2}} }{\|p_{0}\|^{4}},
\end{equation*}
provided that, over $n\leq m$ collisions, the particle's speed does not appreciably change. A more careful computation, carried out in \cite{adblafp} verifies that this is
indeed the leading order contribution to the correlation function.
Hence, we write
\begin{equation}\label{eq:Mstar}
m_{*}(p_{0})=\frac{\|p_{0}\|^{4}}{\overline{
\|\alpha^{\left( 1\right) }\|^{2}} }.
\end{equation}
 To justify the assumption that the particle's speed does not change appreciably during the first $m_*(p_0)$ collisions, recall that, as we have seen,
the energy distribution of the particle approaches a Boltzmann
distribution with a temperature $T$ given by~\eqref{eq:Tweighteddav}; since $E_*$ is assumed to be large, the same
is true for $k_BT$ and hence the bulk of the particles in the limiting equilibrium distribution, which have speeds comparable to the thermal speed $v_{T}=\sqrt{dk_BT}$, are both fast and energetic, in the sense of~\eqref{eq:energetic} and~\eqref{eq:fast}.
The reasoning above then applies if we can argue that during $m_{T}:=m_{*}(v_{T})$ collisions, such particles do not appreciably change their speed. To see this, note that
from~\eqref{eq:deltaEexpansion}, and from that fact
that $\overline{\beta^{(1)}}=0=\overline{\beta^{(3)}}$, we can infer that
\begin{equation}  \label{eq:deltaEexpansionav}
\overline{\Delta E}(p)=\frac{\overline{\beta^{(2)}}}{\|p\|^{2}} +\mathrm{O}\left(\|p\|^{-4}\right)=-\frac{\alpha\tilde\alpha\overline{(\eta'P)^2L_0^2}}{2M\|p\|^{2}} +\mathrm{O}\left(\|p\|^{-4}\right).
\end{equation}
Hence, over a large number $n$ of collisions, an initially fast particle with momentum $p$ will, to leading
order, \emph{decelerate} on average at a rate
\begin{equation*}
\frac{\mathrm{d} \|p\|}{\mathrm{d} n}=\frac{\mathrm{d} \|p\|}{\mathrm{d} E}\
{\overline{\Delta E}}(p)=\frac{\overline{\beta^{\left( 2\right) }}}{\|p\|^{3}%
}.
\end{equation*}
On average, therefore, it will take on the order of
\begin{equation}\label{eq:Nstar}
n_{*}(p)=\frac{\|p\|^{4}}{4|\overline{\beta^{(2)}}|}
\end{equation}
collisions for the particle to slow down to speeds comparable to the critical speed
\begin{equation*}
 v_*=\max\{\sqrt{\alpha E_*^{r'/r}}, \sqrt{E_*^{1-\frac2r}}\}
\end{equation*}
obtained from~\eqref{eq:energetic}-\eqref{eq:fast}.
For particles in equilibrium, which have speeds of the order $v_{T}$,
we find, setting $n_{T}:=n_*(v_{T})$, that
$$
\frac{n_{T}}{m_{T}}=\frac{\|\alpha^{(1)}\|^2}{\overline{(\beta^{(1)})^2}}\sim \frac{\overline{\eta^2}}{\overline{(\eta')^2}}\sim Q^2\sim T^{2/r}.
$$
Consequently, at high temperatures the average time it takes a typical particle to slow down is much longer
than the one it needs to turn, thus justifying
the assumptions underlying the derivation of~\eqref{eq:Mstar}.

We now conclude. In equilibrium, such particles, which will provide the dominant contribution
to the growth of the mean-squared displacement of the ensemble, travel
for roughly $m_{T}$ collisions
before changing direction significantly. After this many collisions the
direction of motion of such a particle will be uncorrelated with its initial
direction. The process then repeats itself. We therefore expect the
large-scale motion
of the particle to be well approximated by the following random walk:
\begin{equation*}
y_{(k+1)m_{T}}=y_{km_{T}} + \ell_{*}m_{{T}} e_{km_{T}}.
\end{equation*}
In other words, over large enough length scales, the particle motion is essentially
the same as if it traveled on a straight line path for
$m_{T}$ collisions, before turning in a random direction and traveling again
along a straight line over the same distance. Thus, in this last equation,
$y_{km_{T}}$ denotes the position of the particle after a certain number $k$
of these larger excursions. With this picture we can then write
\begin{equation*}
\langle y_{km_{T}}^{2}\rangle= k m_{T}^{2} \ell_{*}^{2}.
\end{equation*}
Extrapolating to all values of $n$, we thus find that
$\langle y_{n}^{2}\rangle= n m_{T}^{2} \ell_{*}^{2}$, which implies
$\langle y_{t}^{2}\rangle=  m_{T}\ell_{*} v_{T} t$,
from which we obtain, using \eqref{eq:Mstar}, the diffusion constant
\begin{equation}  \label{eq:D}
D= m_{T}\ell_{*} v_{T}=\frac{\ell_{*}v_{T}^{5}}{%
\overline{\|\alpha^{(1)}\|^{2}}}.
\end{equation}
\begin{figure}[t]
\begin{center}
\includegraphics[height=6.9cm, keepaspectratio=true]{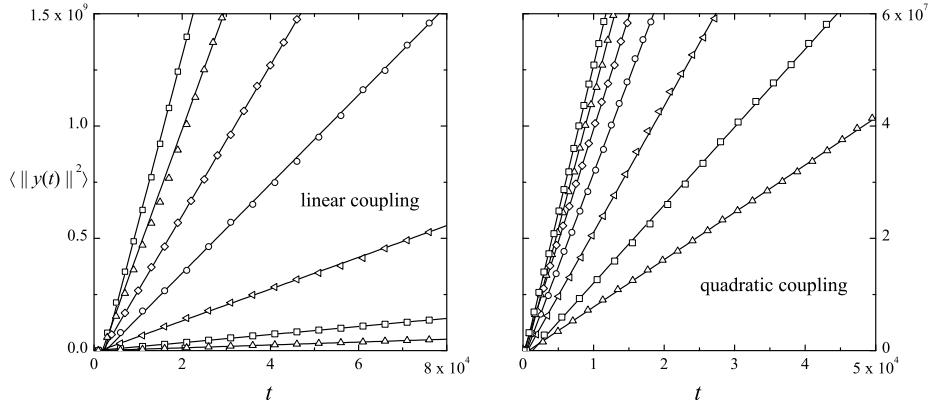}
\end{center}
\caption{Numerically computed mean squared displacements for an ensemble of particles moving through a lattice of
thermally equilibrated harmonic oscillators,
for linear and quadratic couplings as indicated, at thermal energies $k_{\mathrm{B}}T=1,2,5,10,15,20,25$. Steeper slopes in each panel correspond to higher temperatures.}
\label{fig:diffusion}
\end{figure}

The diffusive growth in time of the numerically computed mean-squared displacement of an ensemble of particles moving according to the fully Hamiltonian dynamics~\eqref{eq:eqmotion}, for the model described in Sec.~\ref{s:numerics}, is illustrated in Fig.~\ref{fig:diffusion} for the case of linear and quadratic coupling. In the numerical work
presented in that figure, straight lines indicate linear fits to the numerical data, which are represented by open symbols, and we have again taken $\alpha=0.1$ and $M=1$. The different curves appearing in that figure correspond to temperatures such that $k_{\mathrm{B}}T=1,2,5,10,15,20$ and $25$.

We conclude by providing numerical results that show that this oversimplified argument,
which neglects the variation in the
speed of the particle along its trajectory, captures the essential dependence of
the diffusion constant on the model parameters, in particular on the temperature, and
on the power-laws $|\eta(Q)|\sim |Q|^{r'}$, $U(Q)\sim|Q|^{r}$.
Noting from \eqref{eq:alpha1} that
$$
\overline{\|\alpha^{(1)}\|^{2}}\sim \overline{Q^{2r'}}\sim E_*^{2r'/r}\sim T^{2r'/r},
$$
\begin{figure}
\begin{center}
\includegraphics[height=6.9cm, keepaspectratio=true]{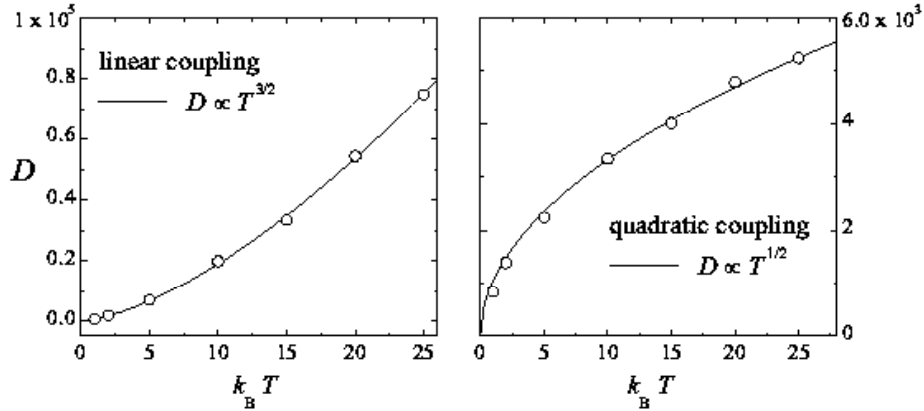}
\end{center}
\caption{Temperature dependence of the diffusion constant $D$ obtained from the data in Fig.~\ref{fig:diffusion} for the case of linear coupling ($r'=1$) and quadratic coupling ($r'=2$).}
\label{fig:diffT}
\end{figure}
one finds from \eqref{eq:D} that
\begin{equation}  \label{eq:DT}
D\sim T^\nu,\qquad\frac12\leq \nu={\frac52-\frac{2r'}{r}}\leq \frac52-\frac2r.
\end{equation}
As mentioned in the introduction, this shows that diffusion is enhanced by reducing the non-linearity in the coupling  (i.e., by lowering $r'$) and by more strongly confining the scatterer degree of freedom (increasing $r$). In particular, for quadratic potentials ($r=2$), our analysis predicts that $D\sim T^{3/2}$ if the coupling is linear, and that $D\sim \sqrt{T}$ if the coupling is quadratic; those predictions are, in fact, as seen in the numerical results presented for these two cases in Fig.~\ref{fig:diffT}, which presents the slopes of the linear fits in
Fig.~\ref{fig:diffusion} as a function of the thermal energy at which they were evaluated.

We conclude this section with a remark on the relation between the
inert Lorentz gas ($\tilde\alpha=0)$, treated in~\cite{adblafp}, and the reactive one
treated here~($\tilde\alpha=\alpha$). Note that $n_*(p)$ in~\eqref{eq:Nstar} behaves as $\tilde\alpha^{-1}$. This is as expected, since it was shown in~\cite{adblafp} that, when $\tilde\alpha=0$, not only does the particle's speed not decrease, it actually increases without bound as $\|p_n\|\sim n^{1/6}$ as a function of the collision number $n$.
But, as shown in~\cite{adblafp}, for initially fast particles of momentum $p$, this stochastic
acceleration begins to manifest itself only after $n_{\mathrm{s}}(p)\sim \|p\|^{6}$ collisions. Thus, the
time scale $n_*(p)\sim \|p\|^4$ over which dynamical friction manifests itself by slowing down fast
particles, and which characterizes the equilibration process, is always much shorter than is
required by the relatively slow process of stochastic acceleration. Thus, in the reactive models
considered here, stochastic acceleration is
completely suppressed and cannot manifest itself before the system equilibrates.
This is a result of the fluctuation-dissipation theorem, which
implies that the frictional component of the force cannot be
independently made small compared to the strength of its fluctuating part.

\section{Discussion}\label{s:discussion}

In this paper we have identified a dynamical mechanism that induces approach to
equilibrium in a classical Hamiltonian system. We have demonstrated numerically and
analytically that as classical particles
scatter repeatedly off the internal degrees of freedom of an array of
independent oscillators with initial conditions drawn from a fixed
invariant distribution, their momentum distribution approaches
a Maxwellian. The effective
temperature that characterizes this distribution is obtained from a generalized
equipartition relation that involves details of the nature of the coupling
between the particle and the scatterers, as well as of the initial
invariant distribution of the scatterers. Since our analytical results are
derived in a high temperature and weak coupling regime, they strictly
speaking only establish the precise behavior of
the equilibrium distribution at large values of the momentum.  The numerical
data presented
in Figs.~\ref{fig:approachCLIN}-\ref{fig:approachMQUAD}, on the other hand,
clearly indicate that the momentum distribution
converges to a Maxwell distribution over the full range of particle speeds,
including very low ones. Identification of the dynamical mechanism responsible for
this agreement with the Maxwell distribution down to the very lowest
speeds is an interesting open problem.

We have also, in our analysis, implicitly focused on the behavior of an ensemble of
particles each of which has an entire medium of scatterers to itself, as it were.
We have, in
particular, not considered effects that might arise in a single system possessing a finite
\emph{density} of moving particles that would, as they equilibrate by exchanging energy with the
scatterers, cause the scatterer distribution itself to evolve (presumably)
towards a limiting distribution.  Indeed, one anticipates that
the state of an ensemble of scatterers would, upon being subjected to repeated collisions
by a sequence of independent
itinerant particles, similarly evolve towards a thermal distribution.
The demonstration of such a result would provide an obvious next step towards a more
complete dynamical demonstration of approach to equilibrium for the dynamical
Lorentz gas, treated as a whole.

The term \emph{dynamical friction} used in the present work was adopted from a series of papers~\cite{chandraI, chandraII, chandraIII} by Chandrasekhar, in which he studies the escape rate of a tracer star from a galactic cluster, the latter being treated as a gas of stars. Chandrasekhar first argues, on general grounds, that if there were no dynamical friction, defined as a deceleration of the star along its direction of motion, galactic clusters could not exist for as long as they do; stars would escape from them too frequently as a result of
stochastically induced fluctuations in their velocities to values exceeding the escape velocity. Assuming the velocity distribution of the stars to be Maxwellian, he then shows that a friction term is indeed produced as a result of successive two-body scattering events between the tracer star and the other stars of the galactic cluster. Finally, he argues that it has a value compatible with the hypothesis of a Maxwellian velocity distribution.

Our analysis and concerns in the present work, although carried out on very different models, are clearly
very similar in spirit to those of Chandrasekhar's.
The main differences between our analysis and that of Chandresekhar can be identified as follows.
Chandrasekhar computes the
average momentum change parallel to the incoming velocity $p$ of the star, per collision, finding
$
\overline{\Delta p_{\parallel}}\sim-{C}/{\|p\|^{3}},
$
where $C>0$ is a constant that depends on the various parameters of the system. The
average momentum change along the direction of motion is therefore negative
and of order $\|p\|^{-3}$, and it is this mechanism that Chandrasekhar identifies
with dynamical friction.

In the notation of the the current work,
this corresponds to a calculation
of the quantity $\overline{e\cdot R(p)}$, and indeed for the current models
we obtain a similar result:
$
\overline{e\cdot R(p)}=
\overline{e\cdot\alpha^{(3)}}\|p\|^{-3}+\dots
$
with
\begin{equation}  \label{eq:thisequation}
\overline{e\cdot\alpha^{(3)}}= \overline{\beta^{(2)}}-\frac12 \overline{{\alpha ^{(1)}\cdot\alpha^{(1)}}} =-\frac{\tilde\alpha\alpha}{2M}
\overline{(\eta')^2}\ \overline{L_0^2}-\frac12\alpha^2\overline\eta^2\overline{\|\nabla L_0\|^2}<0.
\end{equation}
We argue however that strict negativity of $\overline{e\cdot\alpha^{(3)}}$ does not, by itself, allow
one to conclude that the system can maintain, let alone approach, a Maxwellian momentum distribution. Indeed, as
we saw, for that purpose one needs to establish a stronger condition: it
is the coefficient $\overline{\beta^{(2)}}$, that in our analysis clearly governs the effect of dynamical
friction on the particle's speed (or energy), which must be strictly negative.

To emphasize this point we note that, in our models,
\emph{even when the Lorentz gas is inert} so that $\tilde\alpha=0$, the
value of $\overline{e\cdot\alpha^{(3)}}$ is strictly negative,
as it is in Chandresekhar's analysis. But in the inert Lorentz gas there is no dynamical
friction, because $\overline{\beta^{(2)}}=0$ (see~\cite{adblafp}); as a result,
particles in the ensemble do not equilibrate, they undergo an unbounded stochastic acceleration of
the sort that Chandresekhar sought to avoid.  In fact, in our models, and probably
in Chandresekhar's as well, a negative contribution to the average change in the component of
momentum along the direction of motion arises, at least in part, because particles
turn (as we have discussed in Sec. \ref{s:diffusion}), an effect that has nothing to
do with friction or the dissipation of excess energy.

\vskip 1cm

\noindent{\bf Acknowledgments}\ This work was supported by Ministry of Higher Education and Research,
Nord-Pas de Calais Regional Council and FEDER through the Contrat de
Projets Etat Region (CPER) 2007-2013. P.~E.~P.~thanks the Research Centre INRIA-Futurs for its hospitality and support for his stay at SIMPAF, where part of this work was performed.

\appendix

\section{The high momentum expansion of the energy transfer}
\label{s:appendixA}
In this Appendix we  determine
the expansion coefficients $\beta^{(\ell)}$ in (\ref{eq:deltaEexpansion}),
for $\ell=0\dots4$, as well as their means and variances, which are crucial
ingredients of the main results derived in the bulk of the paper. We obtain these
from (\ref{eq:deltadef}) and a perturbative calculation of the solutions to (\ref{eq:onehitequation}) to sufficiently high order.
To this end, it is convenient to introduce a notation in which, for any $k\geq0$
\begin{equation*}
 q(s)=q_{(k)}(s)+\mathrm{O}(s^{k+1}),
\end{equation*}
and similarly for other functions of $s$.
Here, the notation $q_{(k)}(s)$ designates any function of $s$ such that
$q(s)-q_{(k)}(s)=\mathrm{O}(s^{k+1})$; it is, in particular, not unique, an observation that will be helpful in the computations below. Also, since we always have $s\leq t_+$, which is of order $\|p\|^{-1}$, it follows that any term of ${O}(s^{k+1})$ is automatically of ${O}(\|p\|^{k+1})$.
Then, with $q=b-\frac12 e$, we have
\begin{equation*}
\begin{split}
 q_{(3)}(s)=q + ps -\alpha\eta(Q)\nabla K_0(s\|p\|, \|b\|)\|p\|^{-2} \\
-\alpha\frac{P}{M}\eta'(Q)\nabla K_1(s\|p\|, \|b\|)\|p\|^{-3},
\end{split}
\end{equation*}
where, for $b\cdot e=0$, we define, [compare~\eqref{eq:Lk}],
\begin{equation}\label{eq:kn}
\left.
\begin{split}
L_k( \mu, \|b\|)&=\int_0^\mu\rd\lambda\ \sigma(b+(\lambda-\frac12)e)\ \rd\lambda,\\
K_k( \mu, \|b\|)&=\int_0^\mu\rd\lambda\ L_k(\lambda, \|b\|).
\end{split}
\right\}
\end{equation}
Hence $\Delta E$ defined in~\eqref{eq:deltadef} can be written as follows:
\begin{equation}\label{eq:deltaE}
\Delta E= \Delta E_{a} + \Delta E_{b} +\Delta E_{c} +\mathrm{O}(\|p\|^{-5})
\end{equation}
where, with $\Lambda(s)=\eta'(Q(s))\dot Q(s)$,
\begin{eqnarray}
 \Delta E_{a}&=&\alpha\int_0^{t_+} \rd s\ \Lambda(s) \sigma(q+ps)\nonumber\\
\Delta E_{b}&=&-\frac{\alpha^2\eta(Q)}{\|p\|^2}\int_0^{t_+}\rd s\ \Lambda(s) \nabla K_0(s\|p\|, \|b\|)\cdot \nabla\sigma(q+ps)\nonumber\\
\Delta E_{c}&=&-\frac{\alpha^2P\eta'(Q)}{M\|p\|^3}\int_0^{t_+}\rd s\ \Lambda(s) \nabla K_1(s\|p\|,\|b\|)\cdot \nabla\sigma(q+ps).\nonumber
\end{eqnarray}
We need to expand each of these terms to order $\|p\|^{-4}$ included in order to determine $\beta^{(\ell)}, \ell=0\dots4$. In fact, as we have noted in Sec. \ref{s:equilibration}, $\beta^{(0)}=0$ because $t_+$ is of order $\|p\|^{-1}$. Furthermore, since $0\leq s\leq t_+$ and $t_+$ is of the order $\|p\|^{-1}$, to obtain the contributions $\beta^{(\ell)}_a$, $\beta^{(\ell)}_b$, $\beta^{(\ell)}_c$ to $\beta^{(\ell)}$ from $\Delta E_a, \Delta E_b$, and $\Delta E_c$ requires that we expand $\Lambda(s)$, respectively, to order $3$, $1$, and $0$.

We first determine $\beta^{(1))}=\beta^{(1)}_a, \beta^{(2)}=\beta^{(2)}_a$, to which $\Delta E_{b}$ and $\Delta E_{c}$ do not contribute. To expand $\Delta E_a$, we compute, from the equations of motion for $Q(t)$,
\begin{equation*}
 \dot Q_{(3)}(s)=\frac{P}{M}+ S(s),
\end{equation*}
where
\begin{equation}\label{eq:Sdef}
S(s)=-\frac{1}{M}\int_0^s\rd \tau\ \left[U'(Q_{(2)}(\tau))+\tilde\alpha \eta'(Q_{(2)}(\tau))\sigma(q_{(2)}(\tau))\right],
\end{equation}
and hence
\begin{eqnarray}
 Q_{(3)}(s)&=&Q +\frac{P}{M}s -\frac{1}{2M}U'(Q)s^2-\frac{P}{6M^2}U''(Q)s^3\nonumber\\
&\ &\qquad\qquad-\frac{\tilde\alpha}{M\|p\|^2}\eta'(Q)K_0(s\|p\|, \|b\|)\nonumber\\
&\ &\qquad\qquad-\frac{\tilde\alpha P}{M^2\|p\|^3}\eta''(Q)K_1(s\|p\|, \|b\|). \label{eq:Q3}
\end{eqnarray}
From this, one finds, with $L_0$ as defined in~\eqref{eq:kn},
\begin{equation*}
 \dot Q_{(1)}(s)=\frac{P}{M}-\frac{1}{M}U'(Q)s-\frac{\tilde \alpha\eta'(Q)}{M\|p\|}L_0(s\|p\|, \|b\|).
\end{equation*}
Now using the result
\begin{eqnarray}
\Lambda_{(1)}(s) &=&\eta'(Q)\frac{P}{M}-\frac{\eta'(Q)U'(Q)}{M}s\nonumber\\
&\ &\qquad\qquad+\frac{P^2}{M^2}\eta''(Q)s
-\frac{\tilde\alpha\eta'(Q)^2}{M\|p\|}L_0(s\|p\|,\|b\|),\label{eq:Lambda1}
\end{eqnarray}
and
\begin{equation}
\Delta E_{a}=\alpha\int_0^{t_+} \rd s\ \Lambda_{(1)}(s) \sigma(q+ps) +\mathrm{O}(\|p\|^{-3}),
\end{equation}
one easily obtains~\eqref{eq:beta1and2}.

We now turn to
$$
\beta^{(3)}=\beta^{(3)}_a+\beta^{(3)}_b,\quad\mathrm{and}\quad
\beta^{(4)}=\beta^{(4)}_a+\beta^{(4)}_b+\beta^{(4)}_c,
$$
the calculation of which is more involved. Fortunately, in order to establish (\ref{eq:xirw}) only the averages of these quantities, in the sense of~(\ref{eq:overlinedef}), are needed; these are much easier to compute.

As a first step in this calculation we note that, since to lowest order $\Lambda(0)=\eta'(Q)P/M$,
\begin{eqnarray}
\Delta E_{c}&=&-\frac{\alpha^2P^2\eta'(Q)^2}{M^2\|p\|^4}\int_0^{+\infty}\rd \lambda \nabla K_1(\lambda,\|b\|)\cdot \nabla \sigma (q+\lambda e) +\mathrm{O}(\|p\|^{-5}),\nonumber\\
&=&\frac{\beta^{(4)}_c}{\|p\|^4}.\label{eq:beta4c}
\end{eqnarray}
Next, using~\eqref{eq:Lambda1}, we easily determine the leading terms of
\begin{eqnarray*}
\Delta E_{b}&=&-\frac{\alpha^2\eta(Q)}{\|p\|^2}\int_0^{t_+}\rd s\ \Lambda_{(1)}(s) \nabla K_0(s\|p\|, \|b\|)\cdot \nabla\sigma(q+ps)\\
&\ &\qquad\qquad\qquad+\mathrm{O}(\|p\|^{-5})\\
&=&\frac{\beta^{(3)}_b}{\|p\|^3}+\frac{\beta^{(4)}_b}{\|p\|^4} +\mathrm{O}(\|p\|^{-5}),
\end{eqnarray*}
where
\begin{eqnarray}\label{eq:betab4}
 \beta^{(3)}_b&=&-\alpha^2\eta(Q)\eta'(Q)\frac{P}{M}\int_0^{+\infty}\rd\lambda\ \nabla K_0(\lambda, \|b\|)\cdot \nabla\sigma(q+\lambda e)\nonumber\\
 \beta^{(4)}_b&=&-\frac{\alpha^2\eta(Q)}{M^2}\left[P^2\eta''(Q)-M\eta'(Q)U'(Q)\right]\nonumber\\
&\ &\qquad\qquad\qquad\times\int_0^{+\infty}\rd\lambda\ \lambda\nabla K_0(\lambda, \|b\|)\cdot\nabla\sigma(q+\lambda e)\nonumber\\
&\ &-\frac{\alpha^2\tilde\alpha\eta(Q)\eta'(Q)^2}{M}\nonumber\\
&\ &\qquad\qquad\qquad\times\int_0^{+\infty}\rd\lambda\ L_0(\lambda, \|b\|) \nabla K_0(\lambda,\|b\|) \cdot\nabla\sigma(q+\lambda e).\nonumber\\
\end{eqnarray}
Now, to compute $\beta^{(3)}_a$ we need to identify the  $s^2$ contribution to  $\Lambda(s)$. For that purpose, we first write (\ref{eq:Sdef}) as
\begin{equation}\label{eq:Ssplit}
S(s)=S_1(s)+S_2(s)+\mathrm{O}(s^5),
\end{equation}
where
\begin{eqnarray*}
 S_1(s)&=& -\frac1{M}\int_0^s \rd \tau \left[ U'(Q_{(2)}(\tau))+\tilde\alpha \eta'(Q_{(2)}(\tau))\sigma(q+p\tau)\right]\\
S_2(s)&=&\frac{\tilde\alpha\alpha}{M\|p\|^{2}}\eta(Q)\int_0^s\rd\tau\ \eta'(Q_{(2)}(\tau))\nabla K_0(\|p\|\tau, \|b\|)\cdot \nabla\sigma(q+p\tau)\\
&=&\frac{\tilde\alpha\alpha}{M\|p\|^{3}}(\eta\eta')(Q)\int_0^{\|p\|s}\rd\lambda  \nabla K_0(\lambda,\|b\|)\cdot \nabla \sigma(q+\lambda e)+\mathrm{O}(s^4).
\end{eqnarray*}
Because $S_1(s)$ is of order $s$ and $S_2(s)$ is of order $s^3$, it follows that
$$
\Lambda(s)=\frac{P}{M}\eta(Q_{(2)}(s)) + S_1(s)\eta(Q+\frac{P}{M}s)+\mathrm{O}(s^3).
$$
Identifying the $s^2$ term in this expression is now easily done using the relevant terms of (\ref{eq:Q3}); one observes that they are linear or cubic in $P$ and hence of zero average in any stationary distribution $\rho$. In conclusion, since $\beta_c^{(3)}=0$,
\begin{equation}\label{eq:beta3av}
 \overline{\beta^{(3)}}=\overline{\beta_a^{(3)}}+\overline{\beta_b^{(3)}}=0.
\end{equation}

It remains to determine $\overline{\beta^{(4)}}$. Since we have already computed $\beta_b^{(4)}$ and $\beta_c^{(4)}$ [see~\eqref{eq:betab4} and~\eqref{eq:beta4c}], it is sufficient to determine $\overline{\beta_a^{(4)}}$. To simplify the computation, we make the following two observations. First, when $\tilde \alpha=0$ in the equations of motion~\eqref{eq:eqmotion}, the particle moves in a time-dependent potential $\alpha\eta(Q(t))\sigma(q)$ where $Q(t)$ solves the one-dimensional equations of motion of a particle of mass $M$ moving under the influence of the confining potential $U$. As a result, $Q(t)$ is periodic in time, and so, therefore, is the interaction potential $\alpha\eta(Q(t))\sigma(q)$. The corresponding results of~\cite{adblafp}, which were derived under conditions which include this case, can then be applied to this situation. In particular, the coefficients $\overline{\beta^{(i)}}(\tilde\alpha=0)$, for $i=0\dots4$  were computed in~\cite{adblafp} and found to satisfy the equality\footnote{Here and in what follows, whenever a function $f$ depends on $\alpha, \tilde \alpha$, and possibly on other variables such as $\|p\|$ and $\kappa$, we shall write $f(\tilde\alpha=0)$ for the values of the function $f$ on the hyperplane $\tilde \alpha=0$, and $\delta f:=f-f(\tilde\alpha=0)$.}
\begin{equation}
 \overline{\beta^{(4)}}(\tilde\alpha=0)=\frac12(d-3)\overline{(\beta^{(1)})^2}(\tilde\alpha=0).
\end{equation}
Using~\eqref{eq:beta1and2} this yields
\begin{equation}\label{eq:flucdiss}
 \overline{\beta^{(4)}}(\tilde\alpha=0)=\frac12(d-3)\overline{(\beta^{(1)})^2}.
\end{equation}
Hence
\begin{eqnarray}\label{eq:overlinebeta4}
 \overline{\beta^{(4)}}&=&\overline{\beta^{(4)}}(\tilde\alpha=0)+\overline{\delta\beta^{(4)}}\nonumber\\
&=&\frac12(d-3)\overline{(\beta^{(1)})^2}+\overline{\delta\beta_a^{(4)}}+\overline{\delta\beta_b^{(4)}}
\end{eqnarray}
where we have used the result $\delta\beta_c^{(4)}=0$ which follows from~\eqref{eq:beta4c}. Moreover,  $\delta\beta_b^{(4)}$ can be read off from~\eqref{eq:betab4}, i.e.,
\begin{equation}\label{eq:deltabetab4}
\delta\beta_b^{(4)}=-\frac{\alpha^2\tilde\alpha\eta(Q)\eta'(Q)^2}{M}
\int_0^{+\infty}\rd\lambda\ L_0(\lambda,\|b\|) \nabla K_0(\lambda,\|b\|)\cdot\nabla\sigma(q+\lambda e).
\end{equation}
It is therefore sufficient to determine $\overline{\delta\beta_a^{(4)}}$, which turns out to be less difficult than to determine $\overline{\beta_a^{(4)}}$ directly. Indeed, and this is our second observation, all expansion coefficients that we compute, and in particular $\beta_a^{(4)}$, are polynomial in $\alpha$ and $\tilde \alpha$. Thus, to compute $\delta\beta_a^{(4)}$ it suffices to compute only those terms that are of non-vanishing order in $\tilde\alpha$. The required terms turn out to be of first or second order at the most. Taken together, these two observations allow us to compute far fewer terms, and result in a considerable simplification that is ultimately brought about by the fact that~\eqref{eq:flucdiss} is neither obvious nor easily established. To prove it by direct expansion in the present context would, in fact, necessitate combining terms from $\beta_a^{(4)}(\tilde\alpha=0)$, $\beta_a^{(b)}(\tilde\alpha=0)$, and $\beta_c^{(4)}(\tilde\alpha=0)$. A different, more efficient approach is used in~\cite{adblafp}.

To compute $\delta\beta_a^{(4)}$, we start by rewriting $\Delta E_a$ in the form
\begin{equation}\label{eq:deltaEa}
 \Delta E_a=\alpha A(\|p\|, \kappa, \alpha,\tilde \alpha)\frac{P}{M}+\alpha B(\|p\|, \kappa, \alpha,\tilde \alpha)+\mathrm{O}(\|p\|^{-5}),
\end{equation}
where
\begin{equation}\label{eq:A}
A(\|p\|, \kappa, \alpha,\tilde \alpha)=\frac1{\|p\|}\int_0^1\rd\lambda\ \eta'\left(Q_{(3)}\left(\frac{\lambda}{\|p\|}\right)\right)\sigma(q+\lambda e)
\end{equation}
and
\begin{eqnarray}\label{eq:B}
B(\|p\|, \kappa, \alpha,\tilde \alpha)&=&\frac1{\|p\|}\int_0^1\rd\lambda\ \eta'\left(Q_{(3)}\left(\frac{\lambda}{\|p\|}\right)\right)\nonumber\\
&\ &\qquad\qquad\qquad\qquad\times\ S\left(\frac{\lambda}{\|p\|}\right)\sigma(q+\lambda e)\nonumber\\
&=&B_1(\|p\|, \kappa, \alpha,\tilde \alpha)+B_2(\|p\|, \kappa, \alpha,\tilde \alpha)+\mathrm{O}(\|p\|^{-5}),\nonumber\\
&\ &
\end{eqnarray}
with
\begin{eqnarray}
B_1(\|p\|, \kappa, \alpha,\tilde \alpha)&=&\frac1{\|p\|}\int_0^1\rd\lambda\ \eta'\left(Q_{(3)}\left(\frac{\lambda}{\|p\|}\right)\right)\nonumber\\
&\ &\qquad\qquad\times\ S_1\left(\frac{\lambda}{\|p\|}\right)\sigma(q+\lambda e)\label{eq:B1}\\
B_2(\|p\|, \kappa, \alpha,\tilde \alpha)&=&\frac{\alpha\tilde\alpha}{M\|p\|^{4}}(\eta(\eta')^2)(Q)\int_0^1\rd\lambda\ \sigma(q+\lambda e)\int_0^\lambda\rd\lambda'\ \nonumber\\
&\ &\qquad\qquad\qquad\nabla K_0(\lambda', \|b\|) \cdot\nabla\sigma(q+\lambda'e).\label{eq:B2}
\end{eqnarray}
Then, using~\eqref{eq:deltaEa}, we find that\footnote{Here and in what follows, we write $f^{(k)}$ for the coefficient of $\|p\|^{-k}$ in the expansion of a function $f$  in powers of $\|p\|^{-1}$}
\begin{equation}\label{eq:deltabetaa4}
 \delta\beta_a^{(4)}=\delta\Delta E_a^{(4)}=\alpha\delta A^{(4)}(\kappa, \alpha,\tilde \alpha)\frac{P}{M}+\alpha \delta B^{(4)}(\kappa, \alpha,\tilde \alpha).
\end{equation}
We first compute
\begin{eqnarray*}
\delta A(\|p\|, \kappa, \alpha,\tilde\alpha) &=&A(\|p\|, \kappa, \alpha,\tilde\alpha)-A(\|p\|, \kappa, \alpha, 0)\\
&=&\frac{1}{\|p\|}\int_0^1\rd\lambda\
\delta\eta'\left(Q_{(3)}\left(\frac\lambda{\|p\|}\right)\right)\sigma(q+\lambda e)\\
&=&\frac{1}{\|p\|}\int_0^1\rd\lambda\ \eta''\left(Q_{(3)}\left(\frac\lambda{\|p\|},\tilde\alpha=0\right)\right)\\
&\ &\qquad\times\ \delta  Q_{(3)}\left(\frac\lambda{\|p\|}\right)\sigma(q+\lambda e)+\mathrm{O}(\|p\|^{-5}).
\end{eqnarray*}
Using~\eqref{eq:Q3} this becomes
\begin{equation}\label{eq:Aexpansion}
 \begin{split}
\delta A((\|p\|, \kappa, \alpha,\tilde\alpha) =\qquad\qquad\qquad\qquad\qquad\qquad\qquad\qquad\qquad\qquad\qquad\qquad\\
-\frac{\tilde\alpha}{M\|p\|^3}\eta'(Q)\int_0^1\rd\lambda\ \eta''(Q+\frac{\lambda}{\|p\|}\frac{P}{M}) K_0(\lambda, \|b\|)\sigma(q+\lambda e)\qquad\qquad\\
-\frac{\tilde\alpha}{M^2\|p\|^4}P\eta''(Q)^2\int_0^1\rd\lambda\ K_1(\lambda,\|b\|) \sigma(q+\lambda e)+\mathrm{O}(\|p\|^{-5}).
\end{split}
\end{equation}
One then easily finds that
\begin{equation}\label{eq:Afourthorder}
\begin{split}
\delta A^{(4)}=-\frac{\tilde\alpha}{M^2}P\left[\eta'(Q)\eta'''(Q)\int_0^1\rd\lambda\ \lambda K_0(\lambda,\|b\|)\sigma(q+\lambda e)\right.\\
\left.+\eta''(Q)^2\int_0^1\rd\lambda\ K_1(\lambda,\|b\|)\sigma(q+\lambda e)\right].
\end{split}
\end{equation}
Next, we compute the contributions from $\delta B$ in~\eqref{eq:deltaEa}. The term coming from $B_2$ can be read off immediately from~\eqref{eq:B2}:
\begin{eqnarray}\label{eq:deltaB24}
\delta B_2^{(4)}&=&\frac{\alpha\tilde\alpha}{M}(\eta(\eta')^2)(Q)\int_0^1\rd\lambda\ \sigma(q+\lambda e)\int_0^\lambda\rd\lambda'\ \nonumber\\
&\ &\qquad\qquad\quad\qquad\nabla K_0(\lambda', \|b\|) \cdot\nabla\sigma(q+\lambda'e).
\end{eqnarray}
 For $\delta B_1$, we write
\begin{equation*}
 \delta B_1(\|p\|, \kappa, \alpha,\tilde\alpha)=\frac{1}{\|p\|}\int_0^1\rd\lambda\ I(\lambda)\sigma(q+e\lambda),
\end{equation*}
where
\begin{eqnarray*}
 I(\lambda)&=&\delta\left(\eta'(Q_{(3)})\times S_1\right)(\frac{\lambda}{\|p\|})\\
&=&\delta\eta'(Q_{(3)}(\frac{\lambda}{\|p\|}))\times S_1(\frac{\lambda}{\|p\|})+ \eta'(Q_{(3)}(\frac{\lambda}{\|p\|}, \tilde\alpha=0))\times\delta S_1(\frac{\lambda}{\|p\|})\\
&=&J(\lambda)+\hat J(\lambda),
\end{eqnarray*}
so that
\begin{equation}\label{eq:deltaB14}
\delta B_1^{(4)}(\kappa, \alpha,\tilde\alpha)=\int_0^1\rd\lambda\ \left(J^{(3)}(\lambda)+\hat J^{(3)}(\lambda)\right)\sigma(q+\lambda e).
\end{equation}
It follows from~\eqref{eq:Q3} that
\begin{eqnarray*}
J(\lambda)&=&-\frac{\tilde\alpha}{M\|p\|^2}\eta''\left(Q_{(3)}\left(\frac{\lambda}{\|p\|},\tilde\alpha=0\right)\right)
\nonumber\\
&\ &\quad \times\left[\eta'(Q)K_0(\lambda,\|b\|)+\frac{P}{M\|p\|}\eta''(Q)K_1(\lambda,\|b\|)\right] S_1\left(\frac{\lambda}{\|p\|}\right)\nonumber\\
&\ &\qquad\qquad\qquad\qquad\qquad \qquad\qquad+ \mathrm{O}(\|p\|^{-5})\nonumber\\
&=&-\frac{\tilde\alpha}{M\|p\|^2}\eta''(Q)\eta'(Q)K_0(\lambda,\|b\|) S_1\left(\frac{\lambda}{\|p\|}\right)+ \mathrm{O}(\|p\|^{-4}).
\end{eqnarray*}
Using the fact that
$$
S_1\left(\frac{\lambda}{\|p\|}\right)=-\frac{1}{M\|p\|}\left[\lambda U'(Q)+\tilde\alpha\eta'(Q)
L_0(\lambda,\|b\|)\right] + \mathrm{O}(\|p\|^{-2}),
$$
this finally yields
\begin{equation}\label{eq:J3}
 J^{(3)}(\lambda)= \frac{\tilde\alpha}{M^2}\eta''(Q)\eta'(Q)K_0(\lambda,\|b\|)\left[\lambda U'(Q)+\tilde\alpha\eta'(Q)
L_0(\lambda,\|b\|)\right].
\end{equation}
To compute  $\hat J^{(3)}(\lambda)$, we note that, since $S_1(\lambda/\|p\|)$ is of order $\|p\|^{-1}$, we have
$$
\hat J(\lambda)=\eta'\left(Q_{(2)}\left(\frac{\lambda}{\|p\|},\tilde\alpha=0\right)\right)\delta S_1\left(\frac{\lambda}{\|p\|}\right)+\mathrm{O}(\|p\|^{-4}).
$$
Using~\eqref{eq:Q3}, one then finds that
\begin{eqnarray*}
\delta S_1\left(\frac{\lambda}{\|p\|},\tilde\alpha=0\right)&=&
-\frac1M\int_0^{\frac{\lambda}{\|p\|}}\rd\tau\ \left[\left(U'(Q_{(2)}(\tau))-U'(Q_{(2)}(\tau, \tilde\alpha=0)\right)\right.\\
&\ &\qquad\qquad\qquad\qquad\qquad\left.+\tilde\alpha\eta'(Q_{(2)}(\tau))\sigma(q+p\tau)\right]\\
&=&\frac{\tilde\alpha}{M\|p\|^2}\eta'(Q)\int_0^{\frac{\lambda}{\|p\|}}\rd\tau\ U''(Q_{(2)}(\tau,\tilde\alpha=0))K_0(\tau\|p\|,\|b\|)\\
&\ &-\frac{\tilde\alpha}M\int_0^{\frac{\lambda}{\|p\|}}\rd\tau\ \eta'(Q_{(2)}(\tau))\sigma(q+p\tau)+\mathrm{O}(\|p\|^{-4})\\
&=&\frac{\tilde\alpha }{M\|p\|^3}(\eta'U'')(Q)\int_0^{\lambda}\rd\lambda'\ K_0(\lambda',\|b\|)\\
&\ &\qquad-\frac{\tilde\alpha}{M\|p\|}\eta'(Q)L_0(\lambda,\|b\|)\\
&\ &\qquad-\frac{\tilde\alpha}{M\|p\|^2}\eta''(Q)\frac{P}{M}L_1(\lambda,\|b\|)\\
&\ &\qquad\frac{\tilde\alpha }{2M^2\|p\|^3}(U'\eta'')(Q)L_2(\lambda,\|b\|)\\
&\ &\qquad\frac{\tilde\alpha^2}{M^2\|p\|^3}(\eta'\eta'')(Q)\int_0^\lambda\rd\lambda'\ K_0(\lambda',\|b\|)\sigma(q+\lambda' e)\\
&\ &\qquad-\frac{\tilde\alpha}{M^3\|p\|^3}P^2\eta'''(Q)L_2(\lambda,\|b\|)+\mathrm{O}(\|p\|^{-4})\\
&=&\frac{\delta S_1^{(1)}(\lambda)}{\|p\|}
+\frac{\delta S_1^{(2)}(\lambda)}{\|p\|^2}
+\frac{\delta S_1^{(3)}(\lambda)}{\|p\|^3}+\mathrm{O}(\|p\|^4),
\end{eqnarray*}
and
\begin{eqnarray*}
\eta'\left(Q_{(3)}(\frac{\lambda}{\|p\|},\tilde\alpha=0)\right)&=&\eta'(Q)+\eta''(Q)\frac{P}{M}\frac{\lambda}{\|p\|}\\
&\ &\quad -\frac{1}{2M^2}\left[M(U'\eta'')(Q)-P^2\eta'''(Q)\right]\left(\frac{\lambda}{\|p\|}\right)^2\\
&\ &\qquad\qquad\qquad\qquad\qquad\qquad + \mathrm{O}(\|p\|^{-3}),
\end{eqnarray*}
so that
\begin{eqnarray}\label{eq:hatJ3}
\hat J^{(3)}(\lambda)&=&\eta'(Q)\delta S_1^{(3)}+\eta''(Q)\frac{P}{M}\lambda\delta S_1^{(2)}(\lambda)\nonumber\\
&\ &\qquad-\frac{1}{M^2}\left[M(U'\eta'')(Q)-P^2\eta'''(Q)\right]\lambda^2\delta S_1^{(1)}.
\end{eqnarray}
One can now compute $\overline{\beta^{(4)}}$ in~\eqref{eq:overlinebeta4} by using~\eqref{eq:deltabetab4}, \eqref{eq:deltabetaa4}, \eqref{eq:Afourthorder}, \eqref{eq:deltaB14}, \eqref{eq:deltaB24}, \eqref{eq:J3} and \eqref{eq:hatJ3}.
Obviously, a general condensed expression for $\overline{\beta^{(4)}}$ and hence of $\overline\gamma$, defined in~\eqref{eq:overlinegamma}, is not readily obtainable. However, one can easily check that, under the hypotheses of this paper, namely $\alpha=\tilde\alpha$ small, $E_*$ large and $|\eta(Q)|\sim |Q|^{r'}$, $U(Q)\sim |Q|^r$, with $0<r'\leq r$,
$$
\overline\gamma=\frac16(d-2)+\epsilon(\alpha, E_*^{-1}).
$$
To see this, it suffices to examine the various error terms and use the fact that $\overline{|Q|^\nu}\sim E_*^{\nu/r}$ and $\overline{P^2}\sim E_*$, for all $\nu\in\N$.

In the case of linear coupling $\eta(Q)=Q$ and an even confining potential $U(-Q)=U(Q)$, the computation considerably simplifies since one then readily sees that $\overline{\delta\beta_b^{(4)}}=0=\delta A^{(4)}=\delta B_2^{(4)}$, so that
\begin{equation}\label{eq:simplebeta4}
 \overline{\beta^{(4)}}=\frac12(d-3)\overline{\left(\beta^{(1)}\right)^2}+\alpha\overline{\delta B_1^{(4)}}
\end{equation}
Moreover, for this case $J^{(3)}(\lambda)=0$, so that
\begin{equation}
 \overline{\delta B_1^{(4)}}=\int_0^1\rd\lambda\ \hat J^{(3)}(\lambda)=\frac{\tilde\alpha }{M}U''(Q)\int_0^1\rd\lambda\int_0^{\lambda}\rd\lambda'\ K_0(\lambda',\|b\|).
\end{equation}
One then finds from~\eqref{eq:overlinegamma} that
\begin{eqnarray*}
 \overline\gamma&=& \frac16(d-2)+\frac{\overline{\delta \beta_a^{(4)}}}{3\Sigma_1^2}
=\frac16(d-2)+\alpha\frac{\overline{\delta B_1^{(4)}}}{3\Sigma_1^2}\\
&=&\frac16\left[(d-2)+\frac{2\overline{U''}}{{k_\mathrm B}T}\frac{\overline{\int_0^1\rd\lambda\int_0^{\lambda}\rd\lambda'\ K_0(\lambda',\|b\|)\sigma(q+\lambda e) }}{\overline{L_0^2}}\right].
\end{eqnarray*}
Since $U''\sim |Q|^{r-2}$, it follows that $\overline{U''}\sim E_*^{1-\frac2r}<<E_*\sim \overline{P^2}$ so that indeed
the last term is negligible for large $E_*$. More explicitly still, for the model studied numerically in this paper, where $d=2$, $U(Q)=\frac12Q^2$ and $\sigma(q)=\chi(1-2\|q\|)$, one finds after an explicit computation of the last term above that for this particular case
\begin{equation}
6\overline\gamma+1=1+\frac{3\pi}{32k_{\mathrm B}T}.
\end{equation}
This produces in~\eqref{eq:Peq} a density of states factor
$$
\|p\|^{1+\frac{3\pi}{32k_{\mathrm B}T}}
$$
differing by $\|p\|^{\frac{3\pi}{32k_{\mathrm B}T}}$ from its Maxwell-Boltzmann value. Even at the relatively small value of the thermal energy ($k_{\mathrm{B}}T=2$) used in the numerical results of Fig.~\ref{fig:approachCLIN}, a deviation in the tails of the distribution of this magnitude in the power law multiplying the Boltzmann factor would be difficult to observe without considerably better statistics.


\bibliographystyle{alpha}
\bibliography{randombibl}

\begin{thebibliography}{ABLP10}

\bibitem[ABLP10]{adblafp}
B.~Aguer, S.~De Bi\`evre, P.~Lafitte, and P.~E. Parris.
\newblock Classical motion in force fields with short range correlations.
\newblock {\em J. Stat. Phys.}, 138(4):780--814, 2010.

\bibitem[BFS00]{bfs}
V.~Bach, J.~Fr\"ohlich, and I.~Sigal.
\newblock Return to equilibrium.
\newblock {\em J. Math. Phys.}, (41):3985--4060, 2000.

\bibitem[Cha43a]{chandraI}
S.~Chandrasekhar.
\newblock Dynamical friction. {I}. {G}eneral considerations: the coefficient of
  dynamical friction.
\newblock {\em Astrophys. J.}, 97:255--262, 1943.

\bibitem[Cha43b]{chandraII}
S.~Chandrasekhar.
\newblock Dynamical friction. {II}. {T}he rate of escape of stars from clusters
  and the evidence for the operation of dynamical friction.
\newblock {\em Astrophys. J.}, 97:263--273, 1943.

\bibitem[Cha43c]{chandraIII}
S.~Chandrasekhar.
\newblock Dynamical friction. {III}. {A} more exact theory of the rate of
  escape of stars from clusters.
\newblock {\em Astrophys. J.}, 98:54--60, 1943.

\bibitem[DJ03]{dj}
J.~Derezi\'nski and V.~Jak\u{s}i\'c.
\newblock Return to equilibrium for pauli-fierz systems.
\newblock {\em Ann. Henri Poincar\'e}, 4(4):739--793, 2003.

\bibitem[JP98]{jp}
Vojkan Jak{\v{s}}i{\'c} and Claude-Alain Pillet.
\newblock Ergodic properties of classical dissipative systems. {I}.
\newblock {\em Acta Math.}, 181(2):245--282, 1998.

\bibitem[KTH91]{kuboII}
R.~Kubo, M.~Toda, and N.~Hashitsume.
\newblock {\em Statistical physics. {II}}, volume~31 of {\em Springer Series in
  Solid-State Sciences}.
\newblock Springer-Verlag, Berlin, second edition, 1991.
\newblock Nonequilibrium statistical mechanics.

\bibitem[SPB06]{spd}
A.A. Silvius, P.E. Parris, and S.~De Bi\`{e}vre.
\newblock Adiabatic-nonadiabatic transition in the diffusive hamiltonian
  dynamics of a classical holstein polaron.
\newblock {\em Phys. Rev. B}, 73:014304, 2006.

\end{thebibliography}

\end{document}